\documentclass[referee]{aa}
\usepackage{graphicx}
\usepackage{wasysym}

\begin{document}

\title{Motion of dust in mean-motion resonances with planets}

\author{P. P\'{a}stor, J. Kla\v{c}ka, L. K\'{o}mar}

\institute{
   Department of Astronomy, Physics of the Earth, and Meteorology, \\
   Faculty of Mathematics, Physics and Informatics, \\
   Comenius University,
   Mlynsk\'{a} dolina, 842~48 Bratislava, Slovak Republic \\
   e-mail: pavol.pastor@fmph.uniba.sk, klacka@fmph.uniba.sk}

\date{}

\abstract{
Effect of stellar electromagnetic radiation on motion of spherical dust
particle in mean-motion orbital resonances with a planet is investigated.
Planar circular restricted three-body problem with the Poynting-Robertson (P-R)
effect yields monotonous secular evolution of eccentricity when the particle is
trapped in the resonance. Elliptically restricted three-body problem with the
P-R effect enables nonmonotonous secular evolution of eccentricity and the
evolution of eccentricity is qualitatively consistent with the published
results for the complicated case of interaction of electromagnetic radiation
with nonspherical dust grain. Thus, it is sufficient to allow either nonzero
eccentricity of the planet or nonsphericity of the grain and the orbital
evolutions in the resonances are qualitatively equal for the two cases. This
holds both for exterior and interior mean-motion orbital resonances. Evolutions
of longitude of pericenter in the planar circular and elliptical restricted
three-body problems are shown. Our numerical integrations suggest that any
analytic expression for secular time derivative of the particle's longitude of
pericenter does not exist, if a dependence on semi-major axis, eccentricity and
longitude of pericenter is considered (the P-R effect and mean-motion resonance
with the planet in circular orbit is taken into account).

Change of optical properties of the spherical grain with the heliocentric
distance is also considered. The change of the optical properties:
i) does not have any significant influence on secular evolution of
eccentricity, ii) causes that the shift of pericenter is mainly in the same
direction/orientation as the particle motion around the Sun. The statements
hold both for circular and noncircular planetary orbits.
}

\authorrunning{P. P\'{a}stor {\it et al.}}
\titlerunning{Motion of dust in mean-motion resonances with planets}
\maketitle

\section{Introduction}

Orbital motion of dust particles in the zones of mean-motion orbital
resonances with a planet is intensively discussed since the paper
by Jackson and Zook (1989) was published. A body is in resonance with a planet
when the ratio of their mean motions (mean motion $n$ $=$ $2 \pi / T$, where
$T$ is orbital period) is the ratio of two small integers. Besides important
gravitational attraction of the Sun and the planet, moving usually in circular
orbit, also the effect of solar electromagnetic radiation on the particle is
considered. Standardly, dust particle is considered to be spherically symmetric
and, correspondingly, the effect of solar electromagnetic radiation is
considered in the form of the Poynting-Robertson (P-R) effect: e. g., Jackson
and Zook (1989), \v{S}idlichovsk\'{y} and Nesvorn\'{y} (1994), Beaug\'{e} and
Ferraz-Mello (1994), Marzari and Vanzani (1994), Liou and Zook (1995),
Liou {\it et al.} (1995), Liou and Zook (1997). Observations confirming the
existence of dust ring around the Sun in resonant lock with the Earth are
discussed in Brownlee (1994), Dermott {\it et al.} (1994), Reach {\it et al.}
(1995). The paper by Liou and Zook (1997) presents secular orbital evolution
of eccentricity when the particle is trapped in mean-motion orbital resonances
(Liou and Zook 1997 -- Eq. 26; see also Kla\v{c}ka and Kocifaj 2006a -- Sec. 9,
mainly Eq. 83). As a consequence, P-R effect and circular restricted three-body
problem yield monotonous secular evolution of eccentricity of the particle
trapped in a mean-motion orbital resonance with a planet. However,
qualitatively new result was obtained by Kla\v{c}ka {\it et al.} (2005), when
a general interaction of electromagnetic radiation with dust grain was taken
into account. Really, arbitrarily shaped dust grain and its interaction with
solar electromagnetic radiation does not yield, in general, monotonous secular
evolution of eccentricity of the particle trapped in mean-motion resonances
(Kla\v{c}ka {\it et al.} 2005a, 2005b; Kla\v{c}ka and Kocifaj 2006a, 2006b).
Thus, an interesting question arises: Does exist any other generalization of
the standardly used access (P-R effect + planet in circular orbit) which can
produce results qualitatively consistent with nonspherical grains?

We concentrate on orbital evolution of spherical dust particles in mean-motion
orbital resonances, in this paper. We take into account nonzero planetary
eccentricity, also. Besides constant optical properties of the particle, also
dependence of optical properties on heliocentric distance is considered, in
accordance with Kocifaj {\it et al.} (2006). Influence of this effect is
usually ignored in literature. In this paper we concentrate on orbital
evolution of dust grains inside resonances: the initial conditions are given
in the form that the particle starts its motion inside the resonances. Results
of our study can be applied also for the case of orbital evolution of particles
captured in the Earth resonant ring, which is well observed (Brownlee 1994,
Dermott {\it et al.} 1994, Reach {\it et al.} 1995). We compare our results
with the results obtained for nonspherical particles which were published
elsewhere. It is shown that nonzero eccentricity of the planet can mimic
behaviour of nonspherical dust grains for circular planetary orbit.

Attempts in dealing with restricted elliptic three-body problem can be found
also in Gonczi {\it et al.} (1983). However, the authors ignore (see their
Eq. 1) the inertial gravitational term and the nongravitational term represents
only a partial component generated by the action of solar electromagnetic
radiation (they would not receive, e. g., results on orbital evolution
presented by Robertson 1937 and Wyatt and Whipple 1950). Thus, our equation of
motion will be more general, and it will really correspond to physics.
Moreover, we are motivated by the results obtained during the last ten years,
which were not known at the time of Gonczi {\it et al.} (1983).

\section{Mean-motion resonances with planets}

According to the third Kepler's law we have
\begin{eqnarray}\label{1}
a^{3} ~ n^{2} &=& G~M_{\odot} ~ \left ( 1 ~-~ \beta \right ) ~,
\nonumber   \\
a_{P}^{3} ~ n_{P}^{2} &=&
		    G~\left ( M_{\odot} ~+~ m_{P} \right ) ~,
\end{eqnarray}
where $a$, $a_{P}$ are semi-major axes of a particle (characterized by optical
parameter $\beta$) and a planet with mass $m_{P}$, $n$ and $n_{P}$ are mean
motions characterizing revolutions around the Sun of the mass $M_{\odot}$.
The first part of Eq. (1) uses the fact that central Keplerian acceleration is
given by the sum of solar gravitational acceleration and radial component of
radiation pressure acceleration. Eq. (1) yields
\begin{equation}\label{2}
a = \left ( 1 ~-~ \beta \right ) ^{1/3} ~
      \left ( \frac{n_{P}}{n} \right ) ^{2/3} ~
      \left ( 1 ~+~ \frac{m_{P}}{M_{\odot}} \right ) ^{- 1/3} ~
      a_{P} ~.
\end{equation}

A particle is in a mean-motion resonance with a planet
when the ratio of their mean motions is the ratio of two small integers.
If the dust particle is in the resonance with the planet, we can define the
$q-$th order exterior resonance by the relation $n_{P} / n$ $=$ $( p + q ) / p$,
where $p$ and $q$ are integer numbers. Similarily, the $q-$th order interior
resonance is defined by the relation $n_{P} / n$ $=$ $p / ( p + q )$, where
$p$ and $q$ are integer numbers. In terms of orbital periods:
$T / T_{P}$ $=$ $( p + q ) / p$ for exterior ,
$T / T_{P}$ $=$ $p / ( p + q )$ for interior resonance.
On the basis of these definitions and Eq. (2), we can immediately write
\begin{equation}\label{3}
a = \left ( 1 ~-~ \beta \right ) ^{1/3} ~
      \left ( \frac{p+q}{p} \right ) ^{2/3} ~
      \left ( 1 ~+~ \frac{m_{P}}{M_{\odot}} \right ) ^{- 1/3} ~
      a_{P} ~,
\end{equation}
for the semi-major axis of the dust particle in the $q-$th order exterior
resonance with the planet of mass $m_{P}$. Similar relation can be obtained
for the interior resonance.

\section{Equation of motion for spherical dust particle}

Let us consider a spherical dust grain under action of gravitational
forces generated by the Sun and a planet moving around the Sun.
Let the grain is under action of solar electromagnetic radiation, too.
The considered electromagnetic radiation effect is known as the
Poynting-Robertson effect. Equation of motion of the particle is considered in
the form
\begin{eqnarray}\label{4}
\frac{d~ \vec{v}}{d ~t} &=& -~
      \frac{G M_{\odot} \left ( 1 - \beta \right )}{r^{2}} ~
      \vec{e}_{R} ~-~ \beta ~ \frac{G M_{\odot}}{r^{2}} ~ \left (
      \frac{\vec{v} \cdot \vec{e}_{R}}{c} ~ \vec{e}_{R}
      + \frac{\vec{v}}{c} \right ) ~-~
\nonumber \\
& & -~G m_{P} \left \{
    \frac{\vec{r} - \vec{r}_{P}}{| \vec{r} - \vec{r}_{P} |^{3}} ~+~
    \frac{\vec{r}_{P}}{| \vec{r}_{P} |^{3}} \right \} ~,
\nonumber \\
\beta &=& \frac{L_{\odot} \pi R^{2}}{4 \pi ~G M_{\odot} m}
      \frac{\bar{Q}'_{pr}}{c}
      = 7.6 \times 10^{-4} ~\bar{Q}'_{pr}
      \frac{\pi R^{2} \left [ m^{2} \right ]}{m \left [ kg \right ]} ~,
\end{eqnarray}
where $\vec{r}$ and $\vec{r}_{P}$ are position vectors of the particle and the
planet (mass $m_{P}$) with respect to the Sun (mass $M_{\odot}$), $r =$
$| \vec{r} |$, $\vec{e}_{R}$ $=$ $\vec{r} / r$, $\vec{v}$ $=$
$d\vec{r} / dt$ is velocity of the particle, $G$ is the universal
gravitational constant, $c$ is speed of light, $m$ is mass of the particle
($m=4\pi R^{3}\rho$/3 we consider homogenous particles with density $\rho$),
$R$ is radius of the particle, $\bar{Q}'_{pr}$ is the spectrally averaged
efficiency factor for radiation pressure and $L_{\odot}$ is the solar
luminosity. As for the Poynting-Robertson effect, we use equation of motion
derived by Robertson (1937), see also Kla\v{c}ka (1992, 2000, 2004).

In this paper we consider Eq. (4) as the equation of motion in the elliptically
restricted three-body problem together with the P-R effect. To obtain planetary
position in elliptical orbit we solve Kepler equation. Initial conditions of
the planet and the particle always correspond to prograde (counter-clockwise)
motion, in our numerical integrations. Moreover, in some cases, we admit that
optical properties of the particle may change with heliocentric distance due to
the change of temperature. Change of optical properties is described by change
of spectrally averaged efficiency factor for radiation pressure
$\bar{Q}'_{pr}$. We have calculated the value of $\bar{Q}'_{pr}$ for particles
with radii $2$ $\mu$m and $5$ $\mu$m in the range of heliocentric distances
from $0$ to $10$ AU. Relation of $\bar{Q}'_{pr}$ on heliocentric distance is
shown in Fig. 1. Method that we used for calculating spectrally averaged
efficiency factor is taken from Kocifaj {\it et al.} (2006) and the results of
Kla\v{c}ka {\it et al.} (2007).

\begin{figure}
\begin{center}
\includegraphics[height=0.25\textheight]{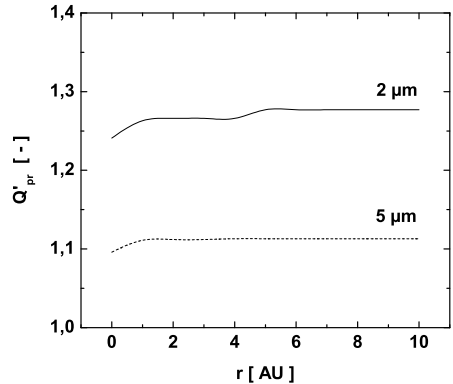}
\end{center}
\caption{Radiation pressure efficiency factor $\bar{Q}'_{pr}$ as a function of
heliocentric distance for compact carbonaceous dust particles with radii
$R=2$ $\mu$m and $R=5$ $\mu$m.}
\label{F1}
\end{figure}

\section{Secular evolution of eccentricity in mean-motion resonances with
planet in circular orbit}

Secular evolution of semi-major axis is characterized by its constant value
when the particle is in a resonance with a planet. What can be said about
secular evolution of the eccentricity of the particle in such a situation? In
this derivation we will suppose that the planet is moving in a circular orbit
around the Sun. In this case, we have a special gravitational problem of three
bodies and small nongravitational forces are also present. The gravitational
problem is know as {\it the circular restricted three-body problem}. At the end
of the 19-th century F. F. Tisserand found a quantity, which does not change
during motion of the third body, whose mass is negligible in comparison to the
masses of planet and the Sun (see, e. g. Brouwer and Clemence 1961). For our
case, we can write the Tisserand's parameter in the form
\begin{equation}\label{5}
C_{T} = \frac{1 - \beta}{a} ~+~2 ~ \sqrt{
    \frac{\left ( 1 - \beta \right ) a \left ( 1 - e^{2} \right )}{a_{P}^{3}}}
    ~ \cos I ~,
\end{equation}
where $e$ is eccentricity of the particle characterized by constant parameter
$\beta$ and $I$ is the inclination of the particle's orbital plane with
reference to the plane of the planet's orbit. We have to stress that
Tisserand's quantity $C_{T}$ does not change only for the special case of the
circular restricted problem of three bodies. However, Eq. (5) enables to find
secular change of the eccentricity of the particle captured in a resonance. For
this purpose, we will consider $I = 0$ in Eq. (5). We can write
\begin{equation}\label{6}
\frac{d C_{T}}{d t} = \frac{\partial C_{T}}{\partial a} ~
		      \left ( \frac{d a}{d t} \right ) _{total} ~+~
		      \frac{\partial C_{T}}{\partial e} ~
		      \left ( \frac{d e}{d t} \right ) _{total} ~,
\end{equation}
for the total time derivative of the Tisserand quantity $C_{T}$ defined by
Eq. (5). However, according to Eq. (4), time derivatives of semi-major axis and
eccentricity of the particle are caused by gravitational perturbations of the
planet (these terms will be denoted by the subscript $G$) and nongravitational
perturbations caused by Poynting-Robertson effect (these terms will be denoted
by the subscript $NG$):
\begin{eqnarray}\label{7}
\left ( \frac{d a}{d t} \right ) _{total} &=&
				    \left ( \frac{d a}{d t} \right ) _{G} ~+~
				    \left ( \frac{d a}{d t} \right ) _{NG} ~,
\nonumber \\
\left ( \frac{d e}{d t} \right ) _{total} &=&
				    \left ( \frac{d e}{d t} \right ) _{G} ~+~
				    \left ( \frac{d e}{d t} \right ) _{NG} ~.
\end{eqnarray}
On the basis of Eqs. (6)-(7) we can write
\begin{eqnarray}\label{8}
\frac{d C_{T}}{d t} &=& \frac{\partial C_{T}}{\partial a} ~ \left \{
			\left ( \frac{d a}{d t} \right ) _{G} ~+~
			\left ( \frac{d a}{d t} \right ) _{NG} \right \} ~+~
\nonumber \\
& &			\frac{\partial C_{T}}{\partial e} ~ \left \{
			\left ( \frac{d e}{d t} \right ) _{G} ~+~
			\left ( \frac{d e}{d t} \right ) _{NG} \right \} ~.
\end{eqnarray}
According to Tisserand, gravitational terms alone do not change the value of
$C_{T}$:
\begin{equation}\label{9}
\frac{\partial C_{T}}{\partial a} ~
		      \left ( \frac{d a}{d t} \right ) _{G} ~+~
		      \frac{\partial C_{T}}{\partial e} ~
		      \left ( \frac{d e}{d t} \right ) _{G} = 0 ~.
\end{equation}
Putting Eq. (9) into Eq. (8):
\begin{equation}\label{10}
\frac{d C_{T}}{d t} = \frac{\partial C_{T}}{\partial a} ~
		      \left ( \frac{d a}{d t} \right ) _{NG}~+~
		      \frac{\partial C_{T}}{\partial e} ~
		      \left ( \frac{d e}{d t} \right ) _{NG} ~.
\end{equation}
If we are interested in secular changes of orbital elements $a$ and $e$, then
the particle's stay in the resonance is characterized by the relation
$\langle da/dt \rangle=0$: for a function $a$ of the property $a(T)=a(0)$ the
relation $\langle da/dt \rangle$ $=$ $(1/T)\int_{0}^{T}(da/dt)dt$ $=$
$(a(T)-a(0))/T$ $=$ $0$ holds. After averaging over period of the resonant
oscillation of semi-major axis Eq. (6) reduces to
\begin{equation}\label{11}
\left \langle \frac{d C_{T}}{d t} \right \rangle =
		      \frac{\partial C_{T}}{\partial e} ~
		      \left \langle \frac{d e}{d t} \right \rangle ~.
\end{equation}
Averaged Eq. (10) and Eq. (11) finally give for the total secular change of the
eccentricity of the particle
\begin{equation}\label{12}
\left \langle \frac{d e}{d t} \right \rangle =
		   \left \langle \frac{d e}{d t} \right \rangle _{NG} ~+~
		   \frac{\partial C_{T} / \partial a}{
		   \partial C_{T} / \partial e } ~
		   \left \langle \frac{d a}{d t} \right \rangle _{NG} ~.
\end{equation}
Calculating partial derivatives of $C_{T}$ defined by Eq. (5), and using
relations for secular changes of the semi-major axis and eccentricity for the
P-R effect (with assuming constant optical properties of particle) (Robertson
1937; Wyatt and Whipple 1950; Kla\v{c}ka 2004)
\begin{eqnarray}\label{13}
\left \langle \frac{d a}{d t} \right \rangle _{NG} = - \beta ~
	\frac{G M_{\odot}}{c} ~\frac{2+3 e^{2}}{a ( 1 - e^{2} )^{3/2}} ~,
\end{eqnarray}
\begin{eqnarray}\label{14}
\left \langle \frac{d e}{d t} \right \rangle _{NG} = - \frac{5}{2} ~ \beta ~
	\frac{G M_{\odot}}{c} ~\frac{e}{a^{2}( 1 - e^{2} )^{1/2}} ~,
\end{eqnarray}
we get for the total secular change of the eccentricity (under the assumption
that Eqs. (13)-(14) hold also for the period of resonant oscillation of the
semi-major axis)
\begin{eqnarray}\label{15}
\left \langle \frac{d e}{dt} \right \rangle &=& \beta ~\frac{G M_{\odot}}{c} ~
		\frac{\left ( 1 - e^{2} \right ) ^{1/2}}{a^{2} ~e} ~ \times ~X ~,
\nonumber \\
X &=& 1 ~-~ \frac{\left ( 1 + 3 e^{2} / 2 \right ) ~ \sqrt{1 - \beta}}{
		\left ( a / a_{P} \right )^{3/2} \left ( 1 - e^{2} \right ) ^{3/2}} ~,
\end{eqnarray}
see Liou and Zook (1997), Kla\v{c}ka and Kocifaj (2006a, 2006b).

Eq. (15) determines secular evolution of eccentricity of the spherical particle
characterized by constant values $\beta$ and $\bar{Q}'_{pr}$, if the particle
is captured into a mean-motion resonance with planet moving in circular orbit.
This equation enables to find detail evolution. If we take some special
mean-motion resonance, we already know the value $n_{P} / n$ and we can
calculate $a / a_{P}$ from Eq. (2). If the initial secular eccentricity $e$ $<$
$e_{\lim}$, where $e_{\lim}$ is given by Eq. (16) below, then the eccentricity
of the particle is an increasing function of time, during the stay of the
particle in the exterior mean-motion resonance. Eccentricity of the particle
can only asymptotically approach to the limiting value $e_{\lim}$ given by the
condition $X = 0$:
\begin{equation}\label{16}
		\frac{p~+~q}{p} = \frac{1 + 3 e_{\lim}^{2} / 2 }{
		( 1 - e_{\lim}^{2} ) ^{3/2}} ~.
\end{equation}
If the initial eccentricity is greater than $e_{\lim}$, then
$\langle de/dt \rangle$ is always negative and the eccentricity of the particle
is a decreasing function of time, during the stay of the particle in the
exterior mean-motion resonance. Eccentricity of the particle can only
asymptotically approach to the limiting value $e_{\lim}$. Characteristic
property of the value $e_{\lim}$ is that $e_{\lim}$ does not depend on $\beta$.
The left part of Fig. 2 depicts evolutions of oscular eccentricity of dust
particle with $\beta=0.01$ in exterior resonance $5/4$ with a planet of mass
equal to the Earth mass, semi-major axis $a_{P}=1$ AU and orbital eccentricity
$e_{P}=0$. The evolutions are calculated from numerical solution of Eq. (4).
The first component of the right-hand side of Eq. (4) is used as a central
acceleration. Asymptotical approach to the limiting value of eccentricity
$e_{\lim} \approx 0.2736$ (given by Eq. 16) can be easily seen. The secular
evolution of eccentricity is always a decreasing function of time for interior
resonances defined by the relation $n/n_{P}$ $=$ $p / ( p + q )$. Evolutions of
oscular eccentricity of dust particle with $\beta=0.01$ in the interior
resonance $2/3$ ($m_{P}$ $= 1$ $m_{Earth}$, $e_{P} = 0$, $a_{P} = 1$ AU) is
shown in the right part of Fig. 2. Evolutions of eccentricities for the given
resonance and $\beta$ are parallel -- the evolutions are shifted along time
axis since Eq. (15) yields the same value of $\langle de/dt \rangle$ for the
same eccentricity $e$. If $\beta=0$ (e. g., an asteroid) we have,
from Eq. (15), $\langle de/dt \rangle=0$.

\begin{figure}
\begin{center}
\begin{minipage}[t]{6cm}
\begin{center}
\includegraphics[height=0.20\textheight]{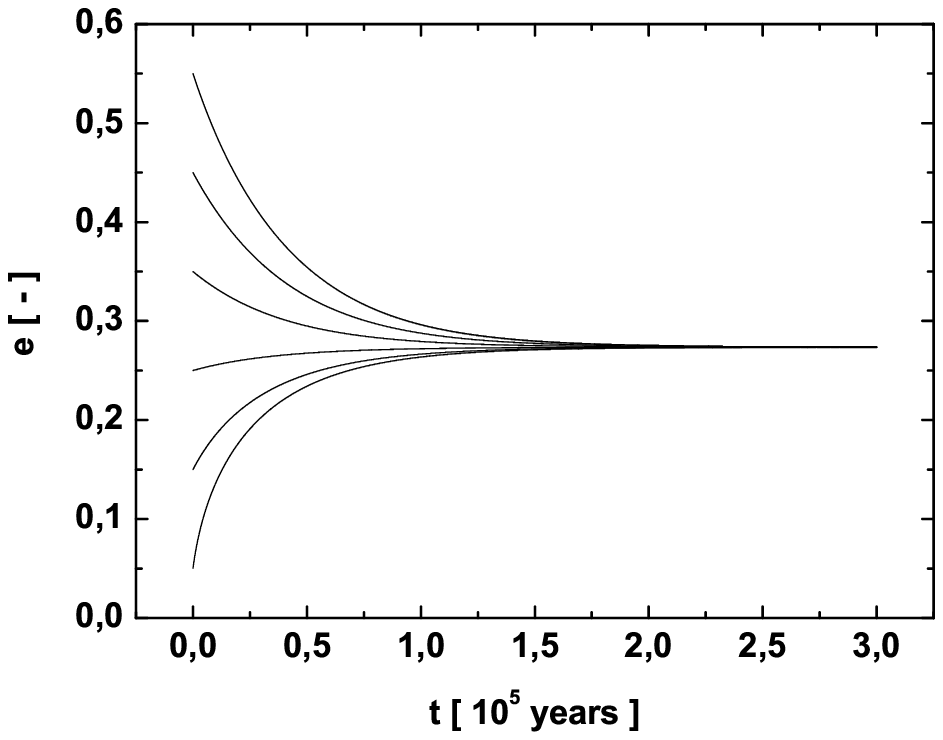}
\end{center}
\end{minipage}
\begin{minipage}[t]{6cm}
\begin{center}
\includegraphics[height=0.20\textheight]{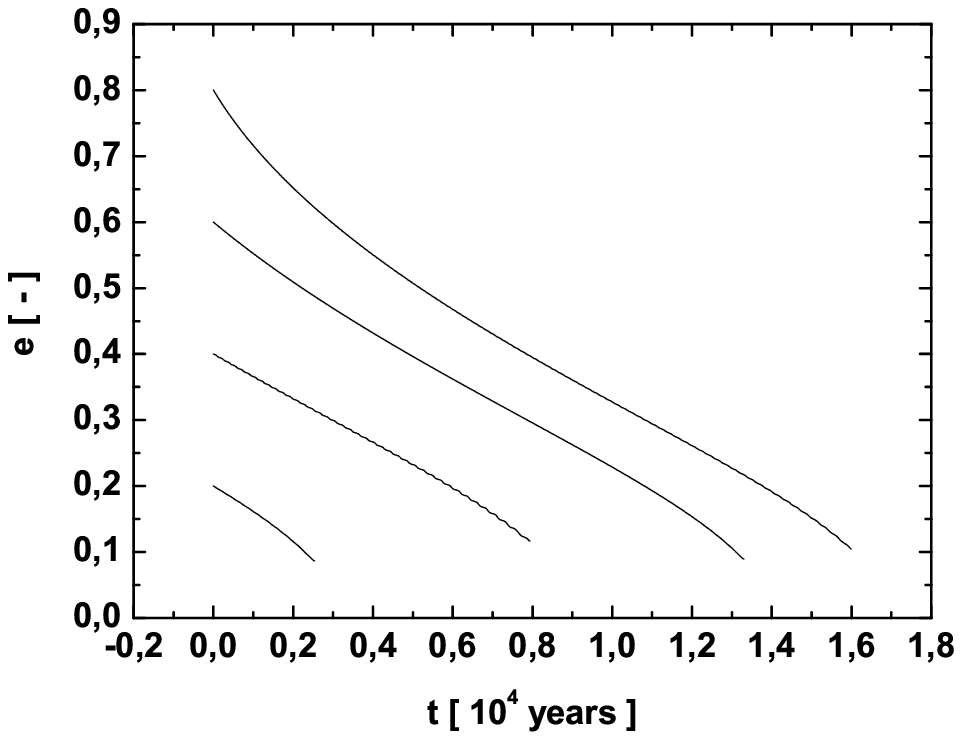}
\end{center}
\end{minipage}
\end{center}
\caption{Evolution of eccentricity of dust particle with $\beta=0.01$
in resonances with a planet of mass equal to the Earth's mass, semi-major axis
$a_{P}=1$ AU and eccentricity $e_{P}=0$. The left part is for $5/4$ exterior
mean-motion resonance. Various evolutions correspond to different initial
values of eccentricity in the resonance. Initial eccentricities are: $0.55$,
$0.45$, $0.35$, $0.25$, $0.15$, $0.05$. The right part of the figure is for
$2/3$ interior mean-motion resonance. Initial eccentricities are: $0.8$, $0.6$,
$0.4$, $0.2$. The evolutions are numerically calculated from Eq. (4) and are
consistent with Eq. (15).}
\label{F2}
\end{figure}

The crucial question emerges: Are the above presented features typical for real
dust particles and real physical situations?

Eq. (15) holds for circular planetary orbit and for the particle with constant
optical properties in a mean-motion resonance with the planet. Evolution of
eccentricity given by Eq. (15) will serve as a reference evolution. It will be
used as a comparison with the evolutions for the cases of non-circular
planetary orbit and non-constant optical properties of the particle.

\section{Secular evolution of longitude of pericenter in mean-motion resonances
with a planet in circular orbit}

Let us assume that a function of the type of Eq. (15) exists for secular
evolution of the particle's longitude of pericenter if the particle is captured
in a mean-motion orbital resonance with a planet in circular orbit in the
planar case. Let us denote this function as $S$. The assumption is that for a
given central star, the planet, the particle and the resonance, the function
$S$ depends on semi-major axis, eccentricity and longitude of pericenter of the
particle:
\begin{equation}\label{17}
\left \langle \frac{d\omega}{dt} \right \rangle=S(a, e,\omega) ~.
\end{equation}
Fig. 3 depicts two evolutions of the orbital elements of the particles with
$\beta=0.01$ in the exterior resonance $5/4$ with the planet of mass equal to
the Earth's mass, semi-major axis $a_{P}=1$ AU and eccentricity $e_{P}=0$.
The particle's initial values of the orbital elements for the first evolution,
depicted by a solid line, are $a \approx 1.1565$ AU (given by Eq. 3), $e=0.2$,
$\omega=90^{\circ}$, and, for the second evolution, depicted by dashed line,
$a \approx 1.1565$ AU, $e=0.2$, $\omega=54^{\circ}$. At the time $t=0$ the
particles are at the perihelia of their orbits. The planet's initial position
is at X-axis (axis, from which the longitude of the pericenter is measured), in
the both integrations. Evolution of the secular eccentricity in Fig. 3 is
consistent with the behavior expected from Eq. (15). We are interested in
secular evolution of the longitude of pericenter. Evolutions of the longitude
of pericenters intersect at the time $t \approx 614.3$ years. Secular values of
the semi-major axes are practically identical, for the both orbits. The same
holds for the eccentricities. However, the values of the
$\langle d\omega/dt \rangle$ differ. This means that the function
$S(a,e,\omega)$ does not exist for the given $a$, $e$ and $\omega$, since
different values of $\langle d\omega/dt \rangle$ occur for the same $a$, $e$
and $\omega$.

Fig. 4 depicts evolution of the orbital elements of the particle with
$\beta=0.01$ in the exterior resonance $5/4$ with the planet of mass equal to
the Earth's mass, semi-major axis $a_{P}=1$ AU and eccentricity $e_{P}=0$.
Particle's initial values are $a \approx 1.1565$ AU, $e=0.2$,
$\omega=0^{\circ}$ (X-axis) for both evolutions. The planetary initial position
is at the X-axis in all integrations. Various particle's initial positions with
respect to the planet yield different orbital evolutions. Various values of
$\langle d\omega/dt \rangle$ exist during a short time interval at the
beginning. This means, as in Fig. 3, that function $S$ does not exist for the
given initial values of $a$, $e$ and $\omega$. Although we have chosen only
particular values of $a$, $e$ and $\omega$, we conjecture that $S(a,e,\omega)$
does not exist for any $a$, $e$ and $\omega$. It is impossible to uniquely
define a function $S$ for various initial conditions $\vec{r}$, $\vec{v}$,
$\vec{r}_{P}$, $\vec{v}_{P}$ yielding a given values of $a$, $e$ and $\omega$.
This does not mean that for the given values of $a$, $e$ and $\omega$ there
cannot exist two various sets of initial conditions leading to the same secular
evolution of the longitude of pericenter. Also in the case $\beta=0$ we have
found evolutions of the longitude of pericenter which have various
$\langle d\omega/dt \rangle$ for the same $a$, $e$ and $\omega$. On the basis
of this result we conjecture that the function $S$ does not exist even in the
case without the P-R effect when only perturbation from the planet is taken
into account.

\begin{figure}
\begin{center}
\begin{minipage}[t]{6cm}
\begin{center}
\includegraphics[height=0.20\textheight]{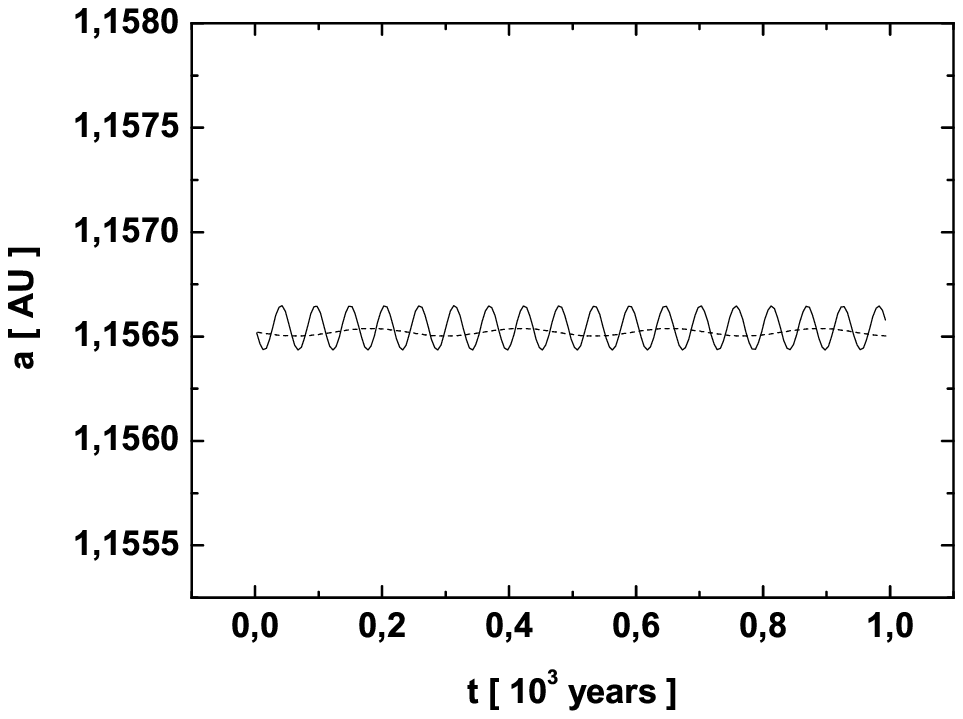}
\end{center}
\end{minipage}
\begin{minipage}[t]{6cm}
\begin{center}
\includegraphics[height=0.20\textheight]{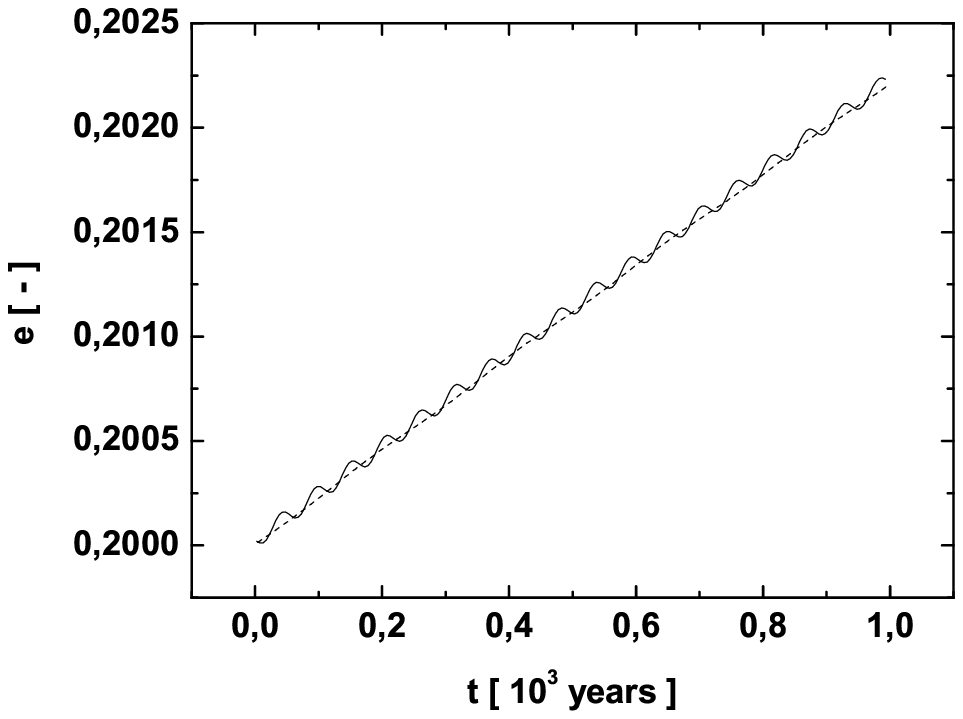}
\end{center}
\end{minipage}
\begin{minipage}[t]{6cm}
\begin{center}
\includegraphics[height=0.20\textheight]{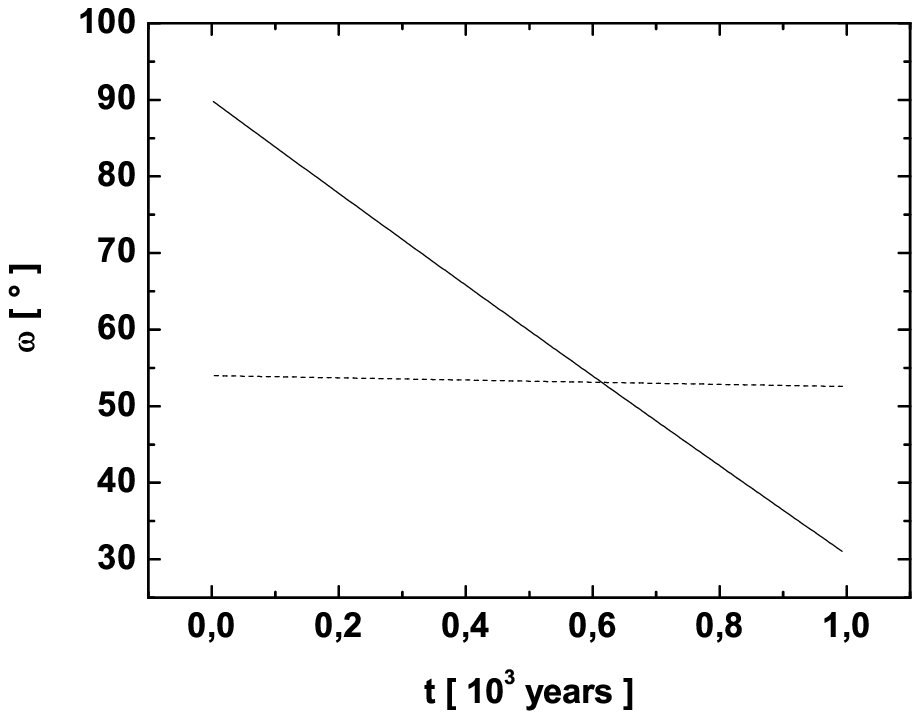}
\end{center}
\end{minipage}
\end{center}
\caption{Two orbital evolutions of dust particle with $\beta=0.01$ captured in
the exterior resonance $5/4$ with a planet of the Earth's mass, semi-major axis
$a_{P}=1$ AU and eccentricity $e_{P}=0$. Initial values of the first evolution
are $a \approx 1.1565$ AU, $e = 0.2$, $\omega = 90^{\circ}$ (solid line), for
the second evolution $a \approx 1.1565$ AU, $e = 0.2$, $\omega = 54^{\circ}$
(dashed line). Evolutions of the longitudes of pericenters intersect at time
$t \approx 614.3$ years.}
\label{F3}
\end{figure}

\begin{figure}
\begin{center}
\begin{minipage}[t]{6cm}
\begin{center}
\includegraphics[height=0.20\textheight]{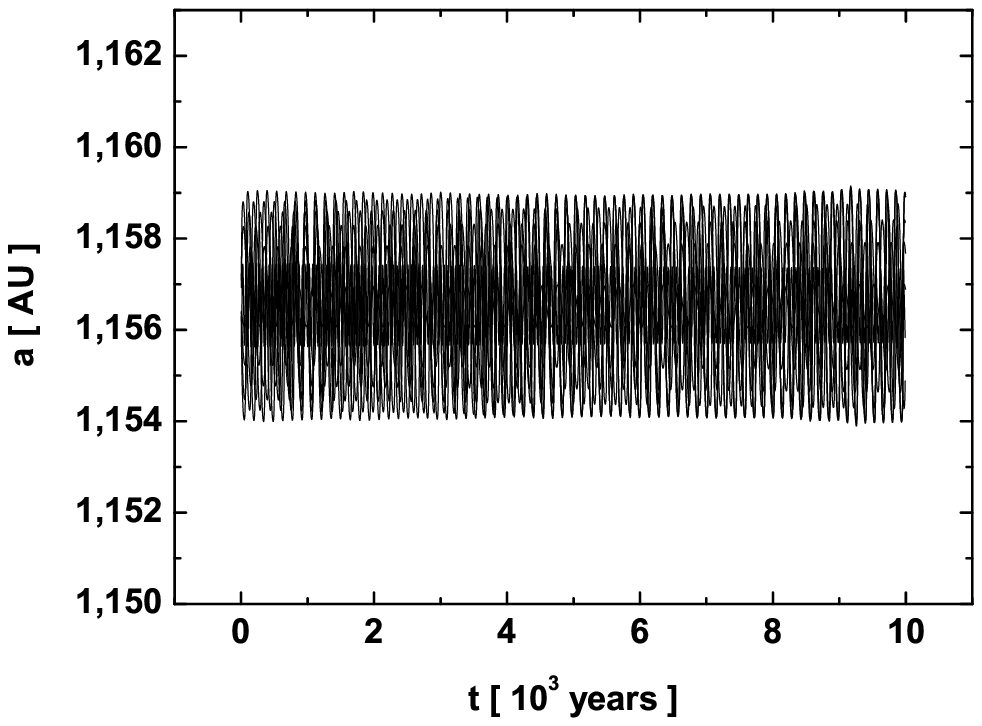}
\end{center}
\end{minipage}
\begin{minipage}[t]{6cm}
\begin{center}
\includegraphics[height=0.20\textheight]{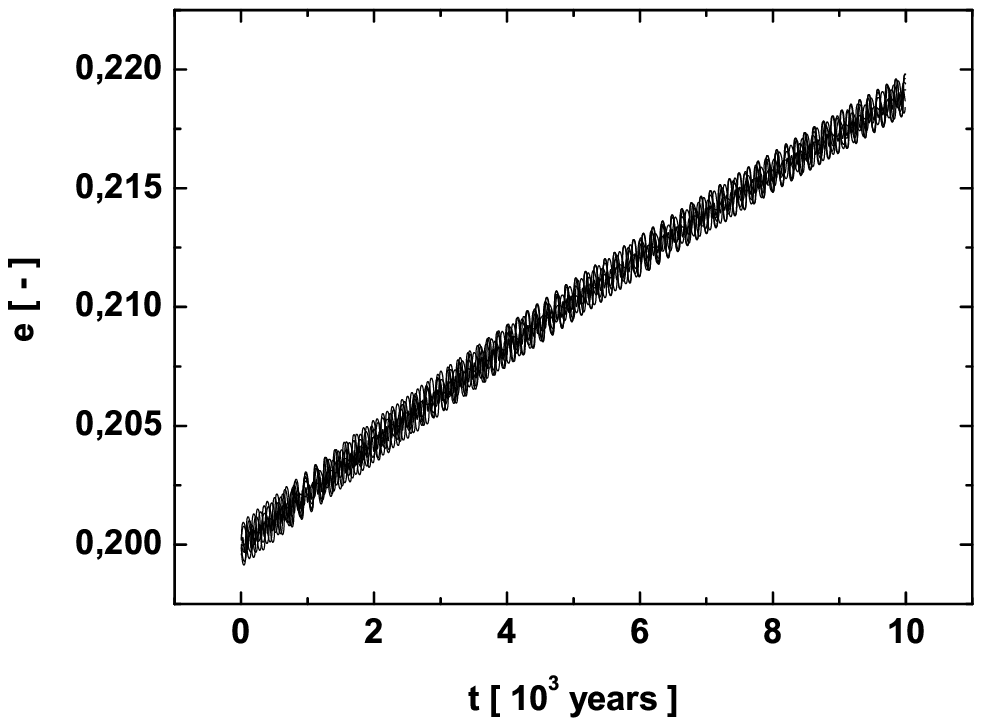}
\end{center}
\end{minipage}
\begin{minipage}[t]{6cm}
\begin{center}
\includegraphics[height=0.20\textheight]{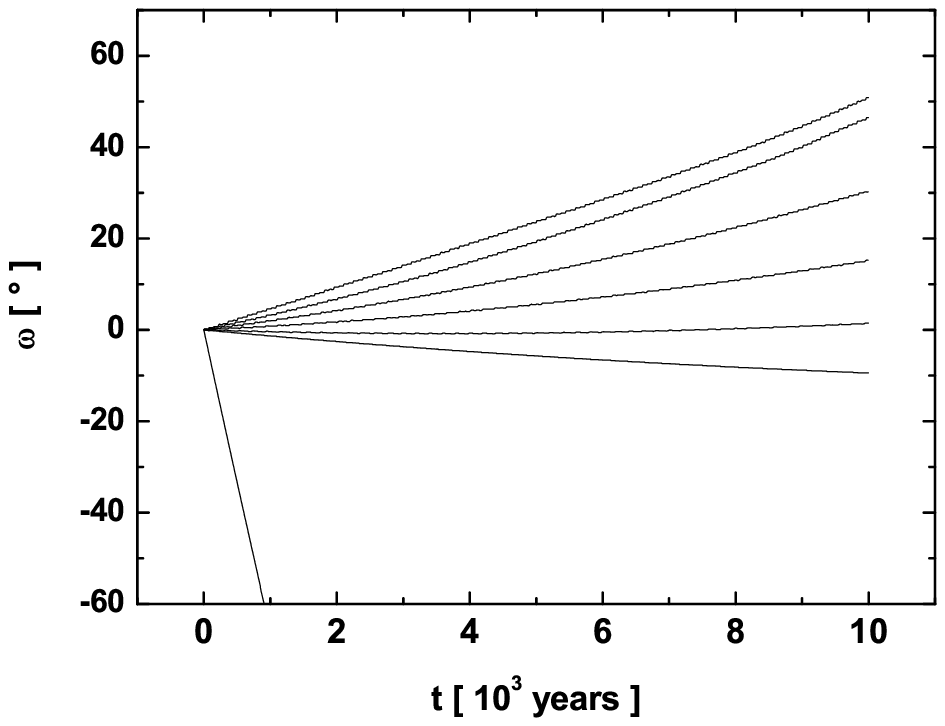}
\end{center}
\end{minipage}
\end{center}
\caption{Orbital evolutions of dust particle with $\beta=0.01$ captured in the
exterior resonance $5/4$ with a planet of the Earth's mass, semi-major axis
$a_{P}=1$ AU and eccentricity $e_{P}=0$. Initial values of all evolutions are
$a \approx 1.1565$ AU, $e = 0.2$, $\omega = 0^{\circ}$. At the beginning, the
particle is localized in the resonance and its initial position varies with
respect to the planet.}
\label{F4}
\end{figure}

\section{Secular evolution of eccentricity and shift of pericenter for
noncircular planetary orbit}

Fig. 5 shows evolution of semi-major axis, eccentricity and longitude of
pericenter of spherical dust grain of constant optical properties with
$\beta=0.01$. The grain is captured in the exterior mean-motion resonance $5/4$
with the planet of mass $m_{P}=1$ $m_{Earth}$ and semi-major axis $a_{P}=1$
AU. Two values of planetary eccentricity are used: $e_{P}=0$ and $e_{P}=0.2$.
Black solid line is used for the case $e_{P}=0.2$, gray and dashed lines for
$e_{P}=0$. Initial eccentricity of the particle is 0.05. Evolution of the
eccentricity for the circular planetary orbit is consistent with Eq. (15):
evolution asymptotically approaches to the limiting value
$e_{\lim} \approx 0.2736$ given by Eq. (16). Evolution of the particle's
eccentricity is a nonmonotonic function of time for the case $e_{P}=0.2$.
Fig. 5 shows that a limiting value of the grain eccentricity may not exist in
the exterior resonance, if $e_{P}>0$.

\begin{figure}
\begin{center}
\begin{minipage}[t]{6cm}
\begin{center}
\includegraphics[height=0.20\textheight]{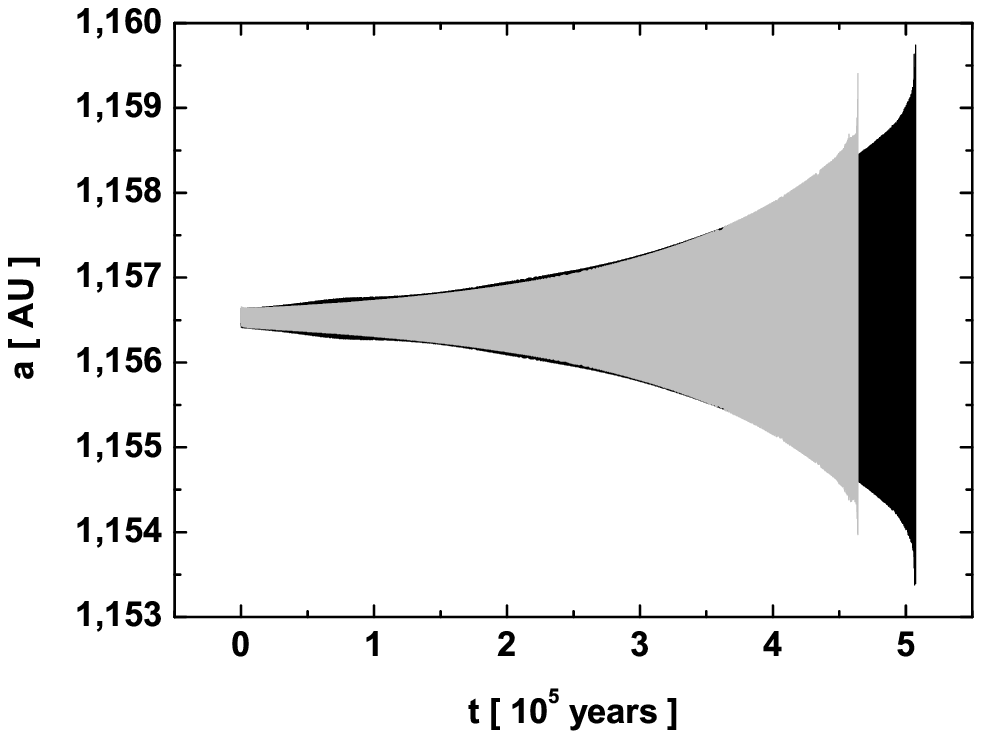}
\end{center}
\end{minipage}
\begin{minipage}[t]{6cm}
\begin{center}
\includegraphics[height=0.20\textheight]{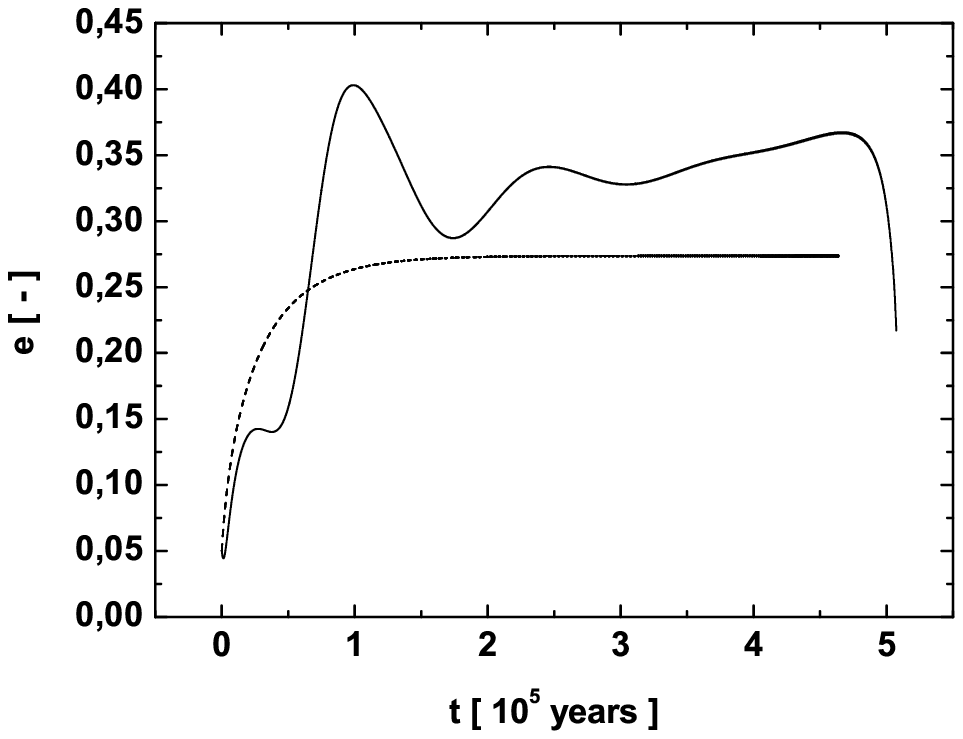}
\end{center}
\end{minipage}
\begin{minipage}[t]{6cm}
\begin{center}
\includegraphics[height=0.20\textheight]{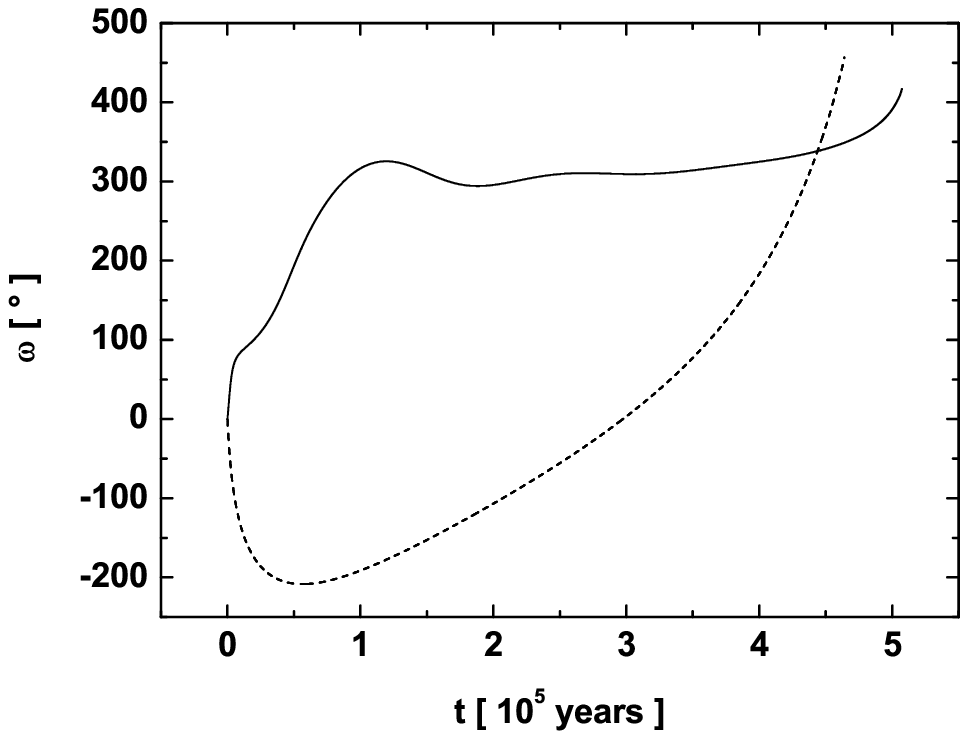}
\end{center}
\end{minipage}
\end{center}
\caption{Orbital evolution of dust particle with constant optical properties.
Particle with $\beta=0.01$ is captured in the exterior resonance $5/4$ with a
planet of the Earth's mass, semi-major axis $a_{P}=1$ AU and eccentricities
$e_{P}=0$ and $e_{P}=0.2$. Initial eccentricity of the particle is $0.05$.
The case $e_{P}=0.2$ corresponds to the black solid line. Gray solid line
corresponds to the evolution of semi-major axis and dashed line to the
evolution of eccentricity and longitude of pericenter. The case $e_{P}=0.2$
does not lead to asymptotic behavior of eccentricity, in contrary to the case
$e_{P}=0$.}
\label{F5}
\end{figure}

Fig. 6 depicts resonant evolution similar to Fig. 5. The main difference is
that the black solid line holds for the eccentricity of the planet orbit
$e_{P}=0.0167$ (eccentricity of the Earth orbit), now. Moreover, initial
conditions for the particle eccentricity are different: the initial value
$e=0.2$ for the case $e_{P}=0.0167$ and $e=0.1845$ for the case $e_{P}=0$. The
motivation was to obtain real oscillations of particle's eccentricity around
the artificial, but analytically solvable (see Eq. 15), case given for
$e_{P} = 0$. Period of the oscillations equals to the period of the shift of
pericenter (period corresponding to the shift in $360^{\circ}$). We have also
found oscillatory evolution of the particle's eccentricity for an initial value
$e$ $>$ $e_{\lim} \approx 0.2736$ for the $5/4$ resonance. The evolution of the
particle's eccentricity oscillates around a curve which asymptotically
decreases to the limiting value $e_{\lim} \approx 0.2736$ (see the left part of
Fig. 2).

\begin{figure}
\begin{center}
\begin{minipage}[t]{6cm}
\begin{center}
\includegraphics[height=0.20\textheight]{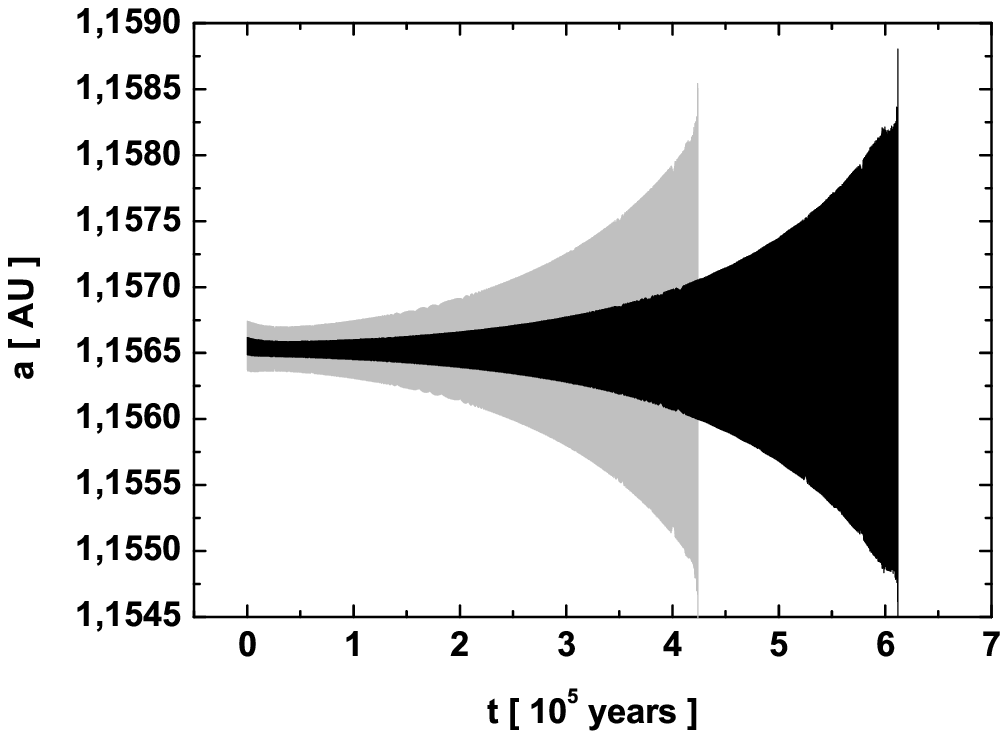}
\end{center}
\end{minipage}
\begin{minipage}[t]{6cm}
\begin{center}
\includegraphics[height=0.20\textheight]{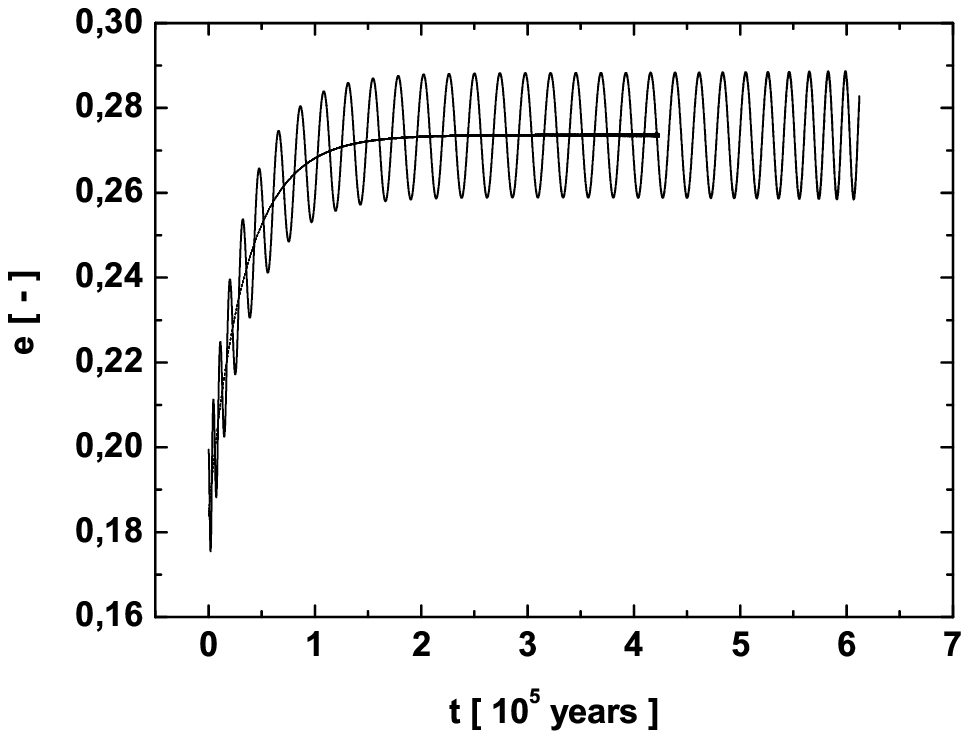}
\end{center}
\end{minipage}
\begin{minipage}[t]{6cm}
\begin{center}
\includegraphics[height=0.20\textheight]{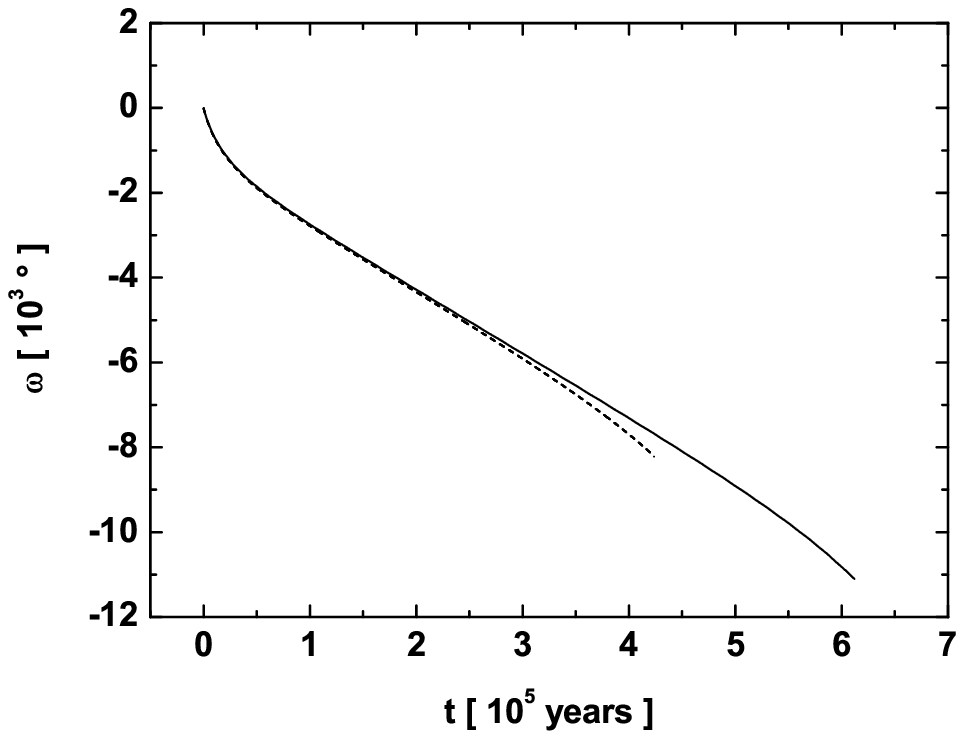}
\end{center}
\end{minipage}
\end{center}
\caption{Orbital evolution of dust particle with constant optical properties.
Particle with $\beta=0.01$ is captured in the exterior resonance $5/4$ with
Earth (semi-major axis $a_{P}=1$ AU and eccentricity $e_{P}=0.0167$) -- black
solid line. Gray color line or dashed line represents evolution of particle
orbital elements in the resonance with the planet in circular orbit. In the
case of the resonance with the Earth, the secular evolution of eccentricity
exhibits oscillations around $e(t)$ given by Eq. (15); period of oscillations
corresponds to the period of the shift of pericenter.}
\label{F6}
\end{figure}

If the shift of pericenter is sufficiently fast, then oscillations in the
evolution of secular eccentricity for exterior resonances exist. Period of this
oscillations corresponds to the period of the shift of pericenter. This
correspondence is caused by libration of conjunctions of the planet and the
particle around aphelion of the particle's orbit. If the aphelion of the
particle's orbit is shifted in $360^{\circ}$, then the planetary orbit has the
same position with respect to the aphelion of the particle orbit (we assume
that secular eccentricity of the particle does not change significantly). This
means that conjunctions of the particle and the planet take place in similar
trajectories and this leads to similar change of the particle's eccentricity.
If the shift of pericenter is slow, then oscillations in secular eccentricity
do not exist. This situation is shown in Fig. 7.

Fig. 7 compares evolutions of orbital elements of dust particle with
$\beta=0.01$ captured in $5/4$ exterior resonance with the planet Earth and an
"artificial Earth" moving in circular orbit. The shift of pericenter in Fig. 7
is much slower than the shift in Fig. 6. As a consequence of the slow shift of
pericenter, the oscillations in evolution of eccentricity are not present in
the first $2.5 \times 10^{5}$ years for the resonance with the Earth. Evolution
of the pericenter during the first $2.5 \times 10^{5}$ is a nonmonotonous
function of time: the initial decreasing function of time (for approximately
$5 \times 10^{4}$ years) is followed by an increasing function.

\begin{figure}
\begin{center}
\begin{minipage}[t]{6cm}
\begin{center}
\includegraphics[height=0.20\textheight]{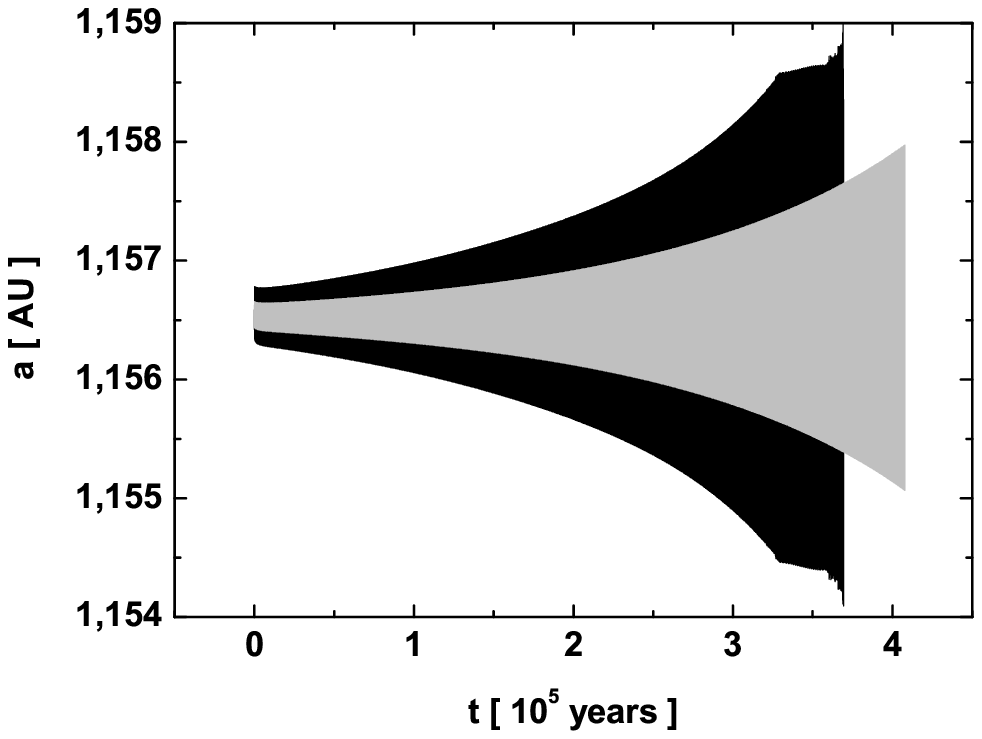}
\end{center}
\end{minipage}
\begin{minipage}[t]{6cm}
\begin{center}
\includegraphics[height=0.20\textheight]{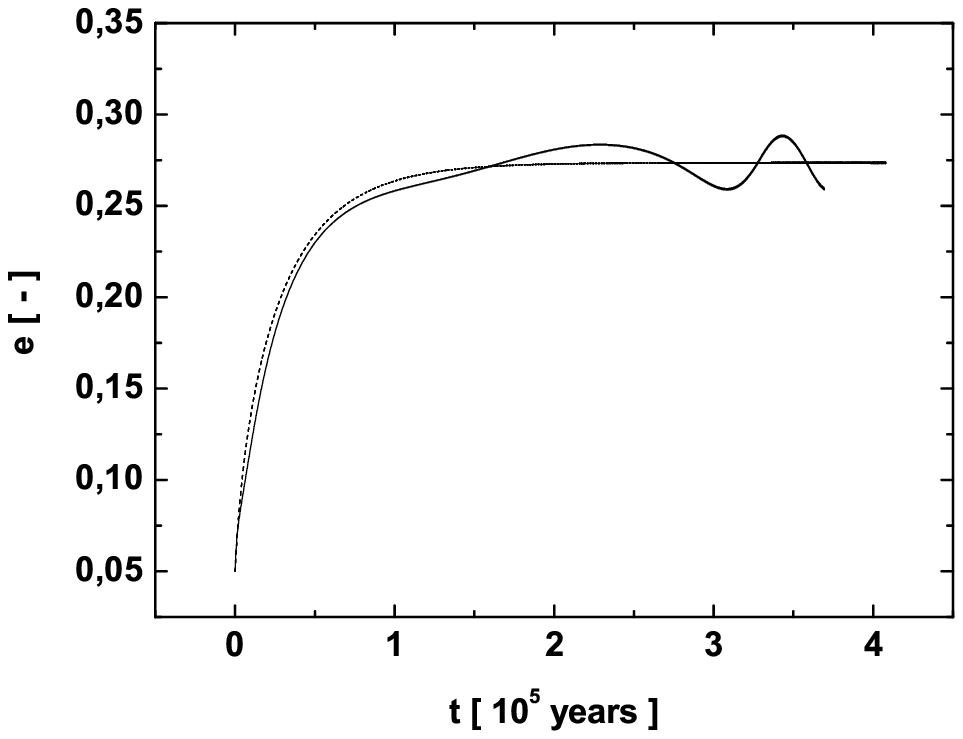}
\end{center}
\end{minipage}
\begin{minipage}[t]{6cm}
\begin{center}
\includegraphics[height=0.20\textheight]{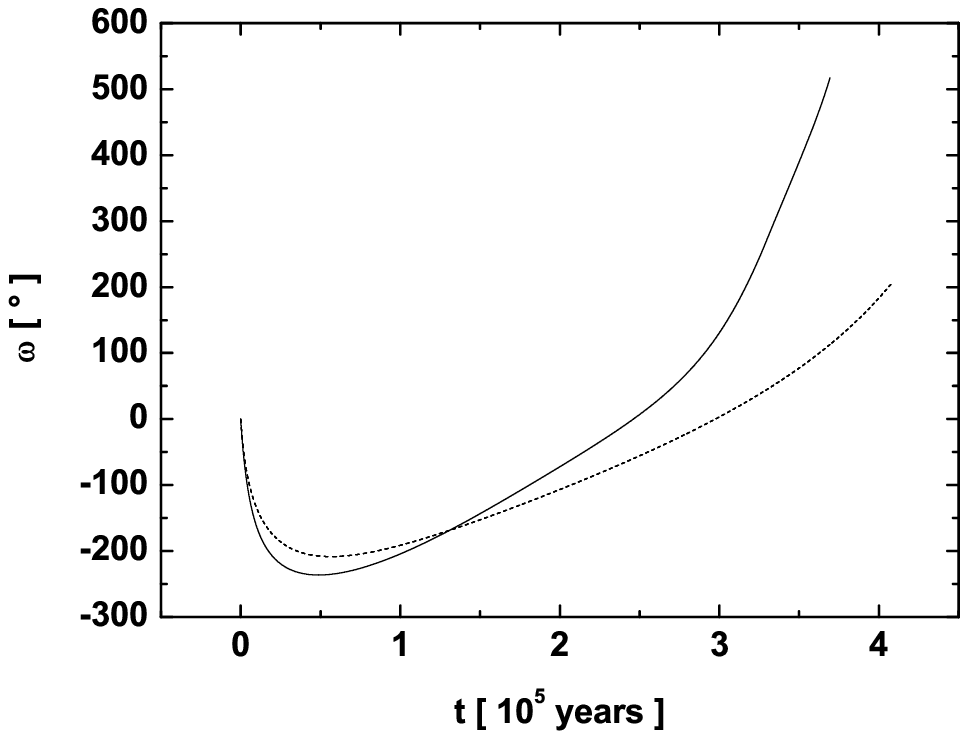}
\end{center}
\end{minipage}
\end{center}
\caption{Orbital evolution of dust particle with constant optical properties.
Particle with $\beta=0.01$ is captured in the exterior resonance $5/4$ with
Earth (semi-major axis $a_{P}=1$ AU and eccentricity $e_{P}=0.0167$) -- black
line. Gray color or dashed curve represents evolution of the orbital elements
of the particle in the resonance with the planet in circular orbit. Initial
conditions differ from those used in Fig. 6. The shift of pericenter is much
slower than the shift in Fig. 6. Evolution of eccentricity exhibits only few
oscillations at the end of the capture when the shift of pericenter is faster.}
\label{F7}
\end{figure}

Our simulations for exterior mean-motion orbital resonances of the first order
show that maximal values of capture times -- for a given $\beta$ and a
resonance $(j+1)/j$ -- for the case $e_{P} > 0$ are in several tens of percent
greater than for the case $e_{P} = 0$. Also the mean capture time for
$e_{p} > 0$ is greater than the mean capture time for $e_{P} = 0$. The same
dependence holds for minimal capture times. This is different from the results
for nonspherical dust grain (compare Kla\v{c}ka {\it et al.} 2005a, 2005b;
Kla\v{c}ka and Kocifaj 2006a, 2006b). The greater aspect ratio and smaller
volume of the nonspherical grain, the more important are nonradial radiation
terms (terms with $\beta_{2}$ and $\beta_{3}$ in Eq. 41 in Kla\v{c}ka 2004) and
the capture times in resonances are shorter. If the distance between the
central star and the planet is greater, then the nonradial radiation terms are
less important than the effect of the planet -- see the term proportional to
$m_{P}$ in Eq. (4) -- and the capture times are greater. These statements for
the nonspherical particles are consistent with the results of Kla\v{c}ka and
Kocifaj (2002), Kla\v{c}ka {\it et al.} (2005a, 2005b), Kla\v{c}ka and Kocifaj
(2006a, 2006b). 

As for the interior mean-motion orbital resonances, we have found that
noncircular planetary orbit can lead to secular increase of the dust grain
eccentricity during the whole capture time; it seems that
$e_{P} > e_{P} (critical) > 0$ is required, where $e_{P} (critical)$ depends on
the type of the resonance, $\beta$ and mass of the planet. This is different
from the case $e_{P} = 0$, when only secular decrease of eccentricity exists
(see Eq. 15). The situation is illustrated in Fig. 8, which holds for resonance
$2/3$, $\beta = 0.01$, $a_{P} = 1$ AU, $e_{P} = 0.2$ and planetary mass equal
to the Earth mass. The obtained result is similar to the evolution of
nonspherical particle with circular planetary orbit (compare Fig. 8 with Fig. 2
in Kla\v{c}ka {\it et al.} 2005b, or, with Fig. 3 in Kla\v{c}ka and Kocifaj
2006b). Fig. 8 shows also very complicated behavior of the shift of pericenter
of the particle. Our simulations show that capture times of particles in the
interior resonances are larger for $e_{P} \ne 0$ than for the case $e_{P} = 0$,
if initial particles eccentricities are small. When eccentricity of the
particle in an interior resonance with the planet in circular orbit decreases
to $0$, the capture is ending because of lack of positive orbital energy given
by the planet (Liou and Zook 1997). However, the eccentricity of the particle
can be an increasing function of time, if the particle is captured in an
interior resonance with the planet moving in elliptical orbit. The particle can
approach to the planet and gain a sufficient amount of positive orbital energy
to prevent a decrease of semi-major axis caused by the P-R effect. This is
explanation of the longer capture times for $e_{P} \ne 0$, if initial particle
eccentricities are small.

\begin{figure}
\begin{center}
\begin{minipage}[t]{6cm}
\begin{center}
\includegraphics[height=0.20\textheight]{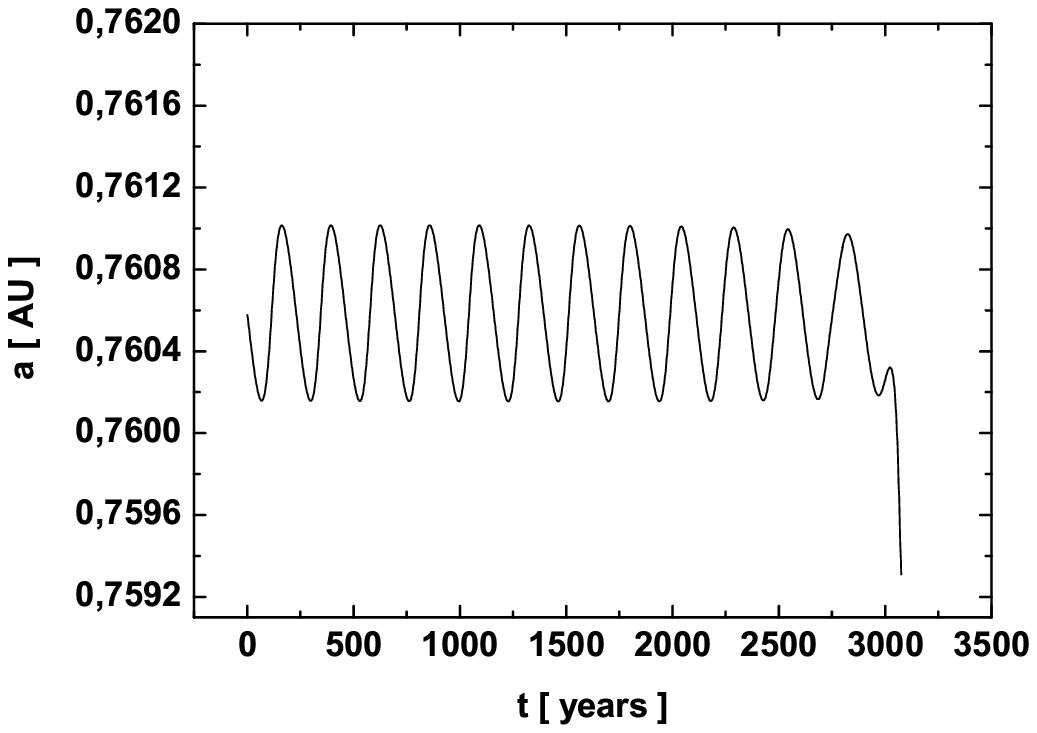}
\end{center}
\end{minipage}
\begin{minipage}[t]{6cm}
\begin{center}
\includegraphics[height=0.20\textheight]{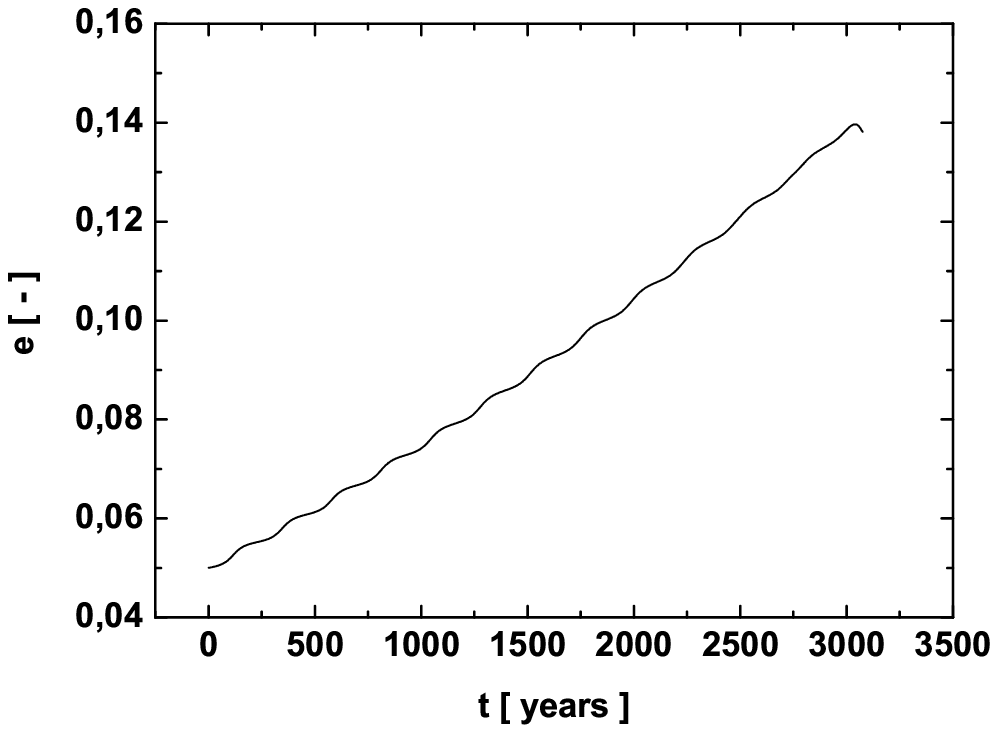}
\end{center}
\end{minipage}
\begin{minipage}[t]{6cm}
\begin{center}
\includegraphics[height=0.20\textheight]{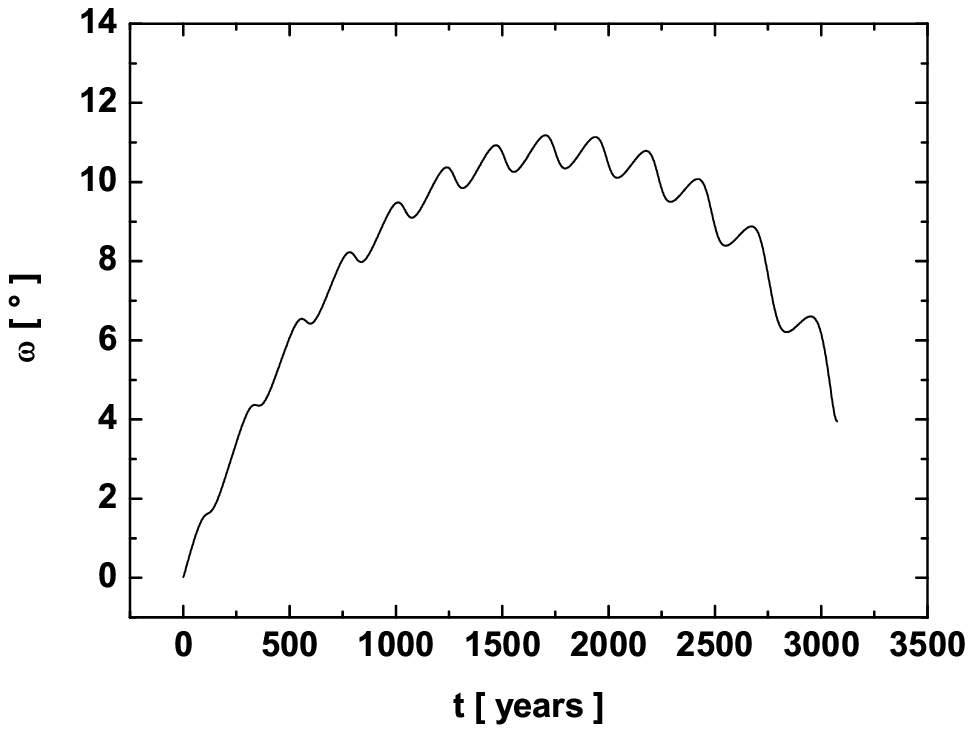}
\end{center}
\end{minipage}
\end{center}
\caption{Secular evolution of orbital elements of dust particle for the
interior resonance $2/3$ with a planet. Spherical particle of constant optical
properties is characterized by $\beta = 0.01$. Mass of the planet equals the
Earth mass, semi-major axis $a_{P} = 1$ AU and planetary eccentricity
$e_{P}= 0.2$. Secular evolution of particle eccentricity is an increasing
function of time during the capture in the resonance. Shift of pericenter is
similar to the case $e_{P} = 0$.}
\label{F8}
\end{figure}

\section{Influence of changing optical properties on evolution of orbital
elements in mean-motion resonances}

We are interested in orbital evolution of a particle with changing optical
properties. As a central acceleration we will use
$-GM_{\odot}(1-\beta_{0})\vec{e}_{R}/r^{2}$, where $\beta_{0}$ is the value of
$\beta$ at the time $t=0$.

We will calculate time derivative of semi-major axis and eccentricity from
perturbation equations of celestial mechanics for Eq. (4) without perturbation
of a planet. We get
\begin{eqnarray}\label{18}
\frac{da_{\beta_{0}}}{dt}&=&
			\frac{2a_{\beta_{0}}e_{\beta_{0}}}{1-e_{\beta_{0}}^{2}}~
			\sqrt{\frac{p_{\beta_{0}}}{GM_{\odot}(1-\beta_{0})}}~(\beta-\beta_{0})
			\frac{GM_{\odot}}{r^{2}}\sin f_{\beta_{0}}-
\nonumber \\
& &    -~\frac{2a_{\beta_{0}}}{1-e_{\beta_{0}}^{2}}~
			\frac{\beta GM_{\odot}}{cr^2}~
			\left [2(e_{\beta_{0}}\sin f_{\beta_{0}})^{2}+
			(1+e_{\beta_{0}}\cos f_{\beta_{0}})^{2} \right ] ~,
\end{eqnarray}
\begin{eqnarray}\label{19}
\frac{de_{\beta_{0}}}{dt}&=&
			\sqrt{\frac{p_{\beta_{0}}}{GM_{\odot}(1-\beta_{0})}}~(\beta-\beta_{0})
			\frac{GM_{\odot}}{r^{2}}\sin f_{\beta_{0}}-
\nonumber \\
& &    -~\frac{\beta GM_{\odot}}{cr^2}~
			\left [e_{\beta_{0}}\sin^{2} f_{\beta_{0}}+
			2(e_{\beta_{0}}+\cos f_{\beta_{0}}) \right ] ~,
\end{eqnarray}
where $a_{\beta_{0}}$, $e_{\beta_{0}}$ and $f_{\beta_{0}}$ are oscular
semi-major axis, eccentricity and true anomaly,
$p_{\beta_{0}}=a_{\beta_{0}}(1-e_{\beta_{0}}^{2})$ is parameter for elliptical
orbit and $r=p_{\beta_{0}}/(1+e_{\beta_{0}}\cos f_{\beta_{0}})$. Secular
evolution of semi-major axis and eccentricity can be obtained by averaging over
one orbital period of the type
\begin{equation}\label{20}
\langle g \rangle = \frac{1}{2\pi a_{\beta_{0}}^{2}\sqrt{1-e_{\beta_{0}}^{2}}}~
\int_{0}^{2\pi}r^{2}g(f_{\beta_{0}})df_{\beta_{0}} ~,
\end{equation}
where $g$ is any quantity. $\beta$ is even function of true anomaly
$f_{\beta_{0}}$. Thus, the average value of the first terms in Eqs. (18)-(19)
is zero. We can write
\begin{eqnarray}\label{21}
\left \langle \frac{da_{\beta_{0}}}{dt} \right \rangle =
&-&      \frac{1}{\pi a_{\beta_{0}}(1-e_{\beta_{0}}^{2})^{3/2}}~
			\frac{GM_{\odot}}{c}\times
\nonumber \\
&\times& \int_{0}^{2\pi}{\beta~
			\left [2(e_{\beta_{0}}\sin f_{\beta_{0}})^{2}+
			(1+e_{\beta_{0}}\cos f_{\beta_{0}})^{2} \right ]}df_{\beta_{0}} ~,
\end{eqnarray}
\begin{eqnarray}\label{22}
\left \langle \frac{de_{\beta_{0}}}{dt} \right \rangle =
&-&      \frac{1}{2\pi a_{\beta_{0}}^{2}\sqrt{1-e_{\beta_{0}}^{2}}}~
			\frac{GM_{\odot}}{c}\times
\nonumber \\
&\times& \int_{0}^{2\pi}{\beta~
			\left [e_{\beta_{0}}\sin^{2} f_{\beta_{0}}+
			2(e_{\beta_{0}}+\cos f_{\beta_{0}}) \right ]}df_{\beta_{0}} ~.
\end{eqnarray}
$\beta$ is approximately constant in Eqs. (21)-(22), see Fig. 1 (for more
details about this approximation see Kla\v{c}ka {\it et al.} 2007). We get
for secular evolution
\begin{eqnarray}\label{23}
\left \langle \frac{da_{\beta_{0}}}{dt} \right \rangle \approx
       -~\beta_{0}\frac{G M_{\odot}}{c}~
			\frac{2+3 e_{\beta_{0}}^{2}}
			{a_{\beta_{0}}(1-e_{\beta_{0}}^{2})^{3/2}} ~,
\end{eqnarray}
\begin{eqnarray}\label{24}
\left \langle \frac{de_{\beta_{0}}}{dt} \right \rangle \approx
       -~\frac{5}{2}~\beta_{0}~\frac{G M_{\odot}}{c}~
			\frac{e_{\beta_{0}}}
			{a_{\beta_{0}}^{2}(1-e_{\beta_{0}}^{2})^{1/2}} ~.
\end{eqnarray}
Eqs. (23)-(24) are identical to Eqs. (13)-(14). Inserting Eqs. (23)-(24) into
Eq. (12) (with the assumption that Eqs. (23)-(24) hold also for period of
resonant oscillation of semi-major axis), we obtain equation which is identical
to Eq. (15). This means that the change of optical properties does not
significantly influence the evolution of secular eccentricity in mean-motion
orbital resonances.

Now, we will derive an expression for secular evolution of the longitude of
pericenter for Eq. (4) without action of a planet. Perturbation equations of
celestial mechanics yield
\begin{eqnarray}\label{25}
\frac{d\omega_{\beta_{0}}}{dt}=
&-&      \frac{1}{e_{\beta_{0}}}~
			\sqrt{\frac{p_{\beta_{0}}}{GM_{\odot}(1-\beta_{0})}}~(\beta-\beta_{0})
			\frac{GM_{\odot}}{r^{2}}\cos f_{\beta_{0}}+
\nonumber \\
&+&~     \frac{1}{e_{\beta_{0}}}\beta
			\frac{GM_{\odot}}{cr^2}~
			\sin f_{\beta_{0}}(e_{\beta_{0}}\cos f_{\beta_{0}}-2) ~.
\end{eqnarray}
After averaging, using also Eq. (20), we have
\begin{eqnarray}\label{26}
\left \langle \frac{d\omega_{\beta_{0}}}{dt} \right \rangle =
&-&      \frac{GM_{\odot}}
			{2\pi a_{\beta_{0}}^{2}e_{\beta_{0}}\sqrt{1-e_{\beta_{0}}^{2}}}~
			\sqrt{\frac{p_{\beta_{0}}}{GM_{\odot}(1-\beta_{0})}}\times
\nonumber \\
&\times& \int_{0}^{2\pi}\beta\cos f_{\beta_{0}}df_{\beta_{0}} ~.
\end{eqnarray}
Since $\beta$ is an even function of true anomaly $f_{\beta_{0}}$, the second
term in the right-hand side of Eq. (25) equals zero, after averaging. Also the
average value of the term proportional to $\beta_{0}$ is zero. The shift of
pericenter caused by the change of optical properties is in the same direction
as the particle orbits the Sun, since $\beta$ is an increasing function of
heliocentric distance (see Fig. 1).

Fig. 9 depicts secular evolution of orbital elements of the particle with
changing optical properties. Solid black line is used for spherical particle
with radius $R=5$ $\mu$m and density $\rho=2$ g/cm$^{3}$ captured in $4/3$
exterior resonance with the Earth. Gray line or dashed line is used for
evolutions of the particle captured in the resonance with the planet of mass
equal to the Earth mass, semi-major axis $a_{P}=1$ AU and eccentricity
$e_{P}=0$. Evolution of eccentricity in Fig. 9 is similar to evolution of
eccentricity in Fig. 6. Evolution of the eccentricity for the case $e_{P} = 0$
asymptotically approaches to the limiting value $e_{\lim} \approx 0.3108$ given
by Eq. (16). Evolution of eccentricity for the resonance with the Earth
exhibits oscillations around the evolution for the resonance with circular
planetary orbit. Period of oscillations corresponds to the period of the shift
of pericenter. Evolution of the longitude of pericenter exhibits an increase in
time. This is the difference from the behavior presented in Fig. 6. Behavior of
the longitude of pericenter in Fig. 9 is caused, probably, by the influence of
the changing optical properties, since the motion of the particle is prograde.

\begin{figure}
\begin{center}
\begin{minipage}[t]{6cm}
\begin{center}
\includegraphics[height=0.20\textheight]{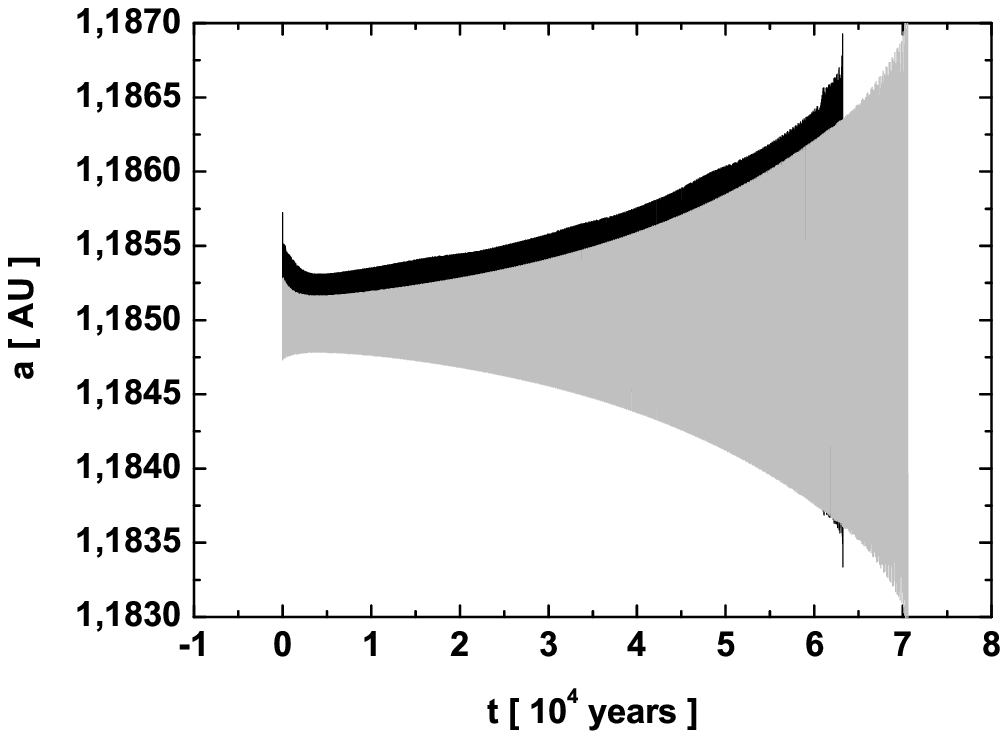}
\end{center}
\end{minipage}
\begin{minipage}[t]{6cm}
\begin{center}
\includegraphics[height=0.20\textheight]{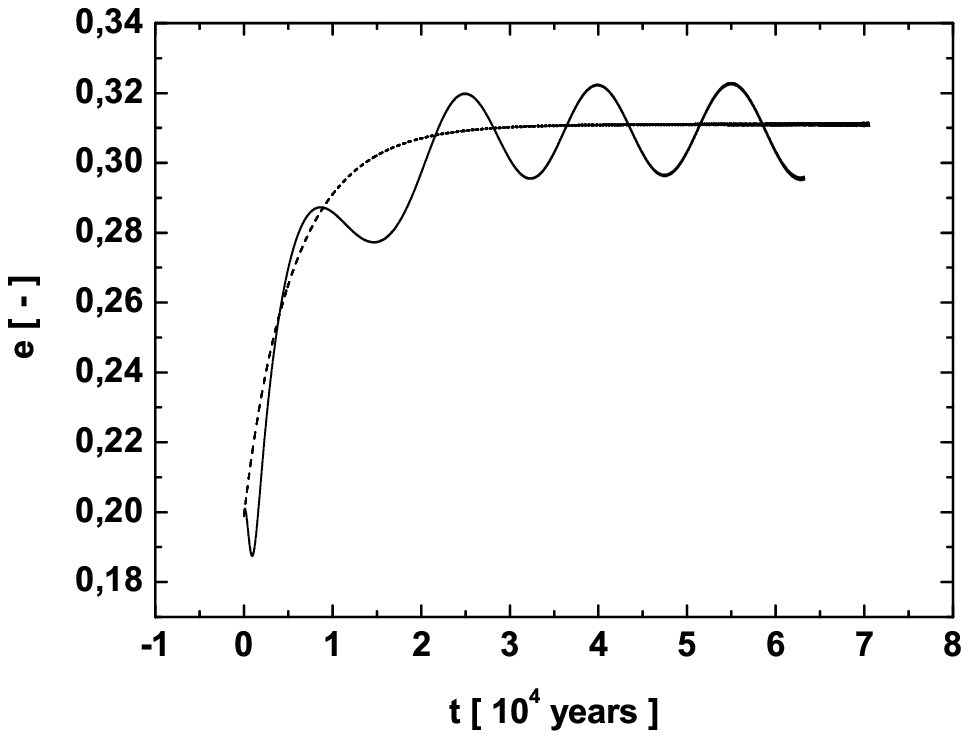}
\end{center}
\end{minipage}
\begin{minipage}[t]{6cm}
\begin{center}
\includegraphics[height=0.20\textheight]{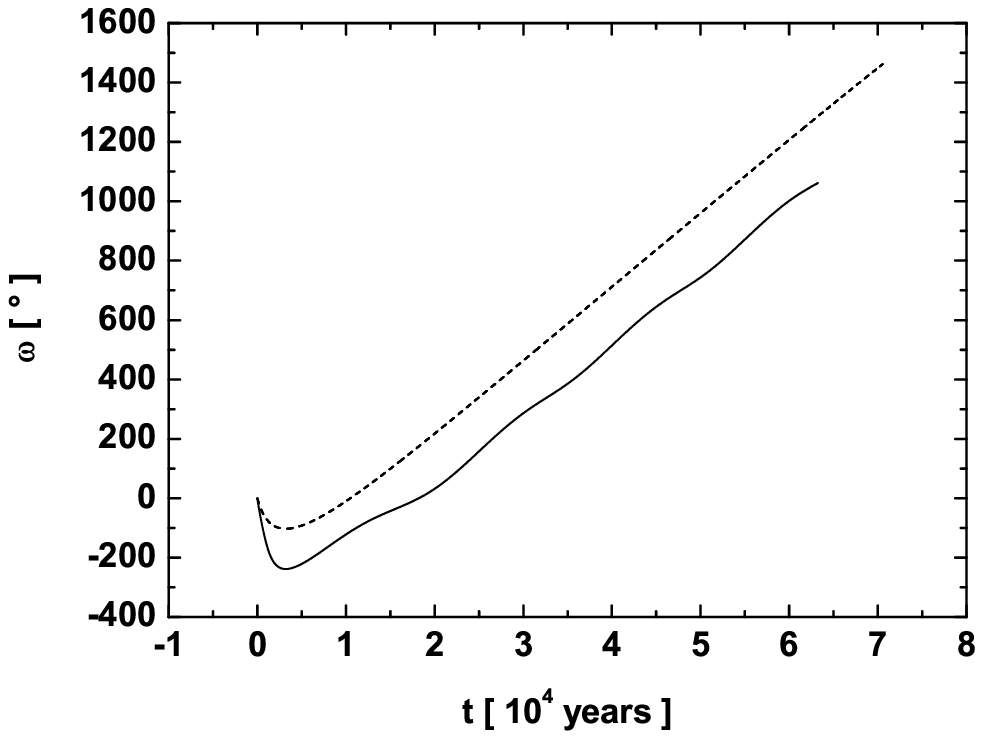}
\end{center}
\end{minipage}
\end{center}
\caption{Orbital evolution of dust particle with optical properties varying
with heliocentric distance; radius of the particle $R=5$ $\mu$m, density
$\rho=2$ g/cm$^{3}$. The particle is captured in the exterior resonance $4/3$
with Earth (semi-major axis $a_{P}=1$ AU and eccentricity $e_{P}=0.0167$)
-- solid black line. Gray color or dashed curve represents orbital evolution of
the particle in the resonance with the planet in circular orbit ($e_{P}=0$). In
the case of the resonance with the Earth, the evolution of eccentricity shows
oscillations around $e(t)$ given by Eq. (15), with period of the shift of
pericenter. Greater part of the shift of pericenter exhibits an increase in
time and this differs from the behavior presented in Fig. 6.}
\label{F9}
\end{figure}

Fig. 10 compares evolution of a particle of constant optical properties with
the evolution of another particle of changing optical properties. Both
particles are characterized by the same initial conditions, as for the orbital
elements and position with respect to the planet. Mass of the planet is equal
to the mass of the Earth, semi-major axis $a_{P}=1$ AU and eccentricity
$e_{P}=0$. The particles are captured in $3/2$ exterior resonance with the
planet. Fig. 10 presents that the shift of pericenter is in positive
direction/orientation, for the case of changing optical properties of the
particle. This is in coincidence with Eq. (26), since the motion is prograde.
The shift of pericenter is in negative direction for particle with constant
optical properties.

\begin{figure}
\begin{center}
\begin{minipage}[t]{6cm}
\begin{center}
\includegraphics[height=0.20\textheight]{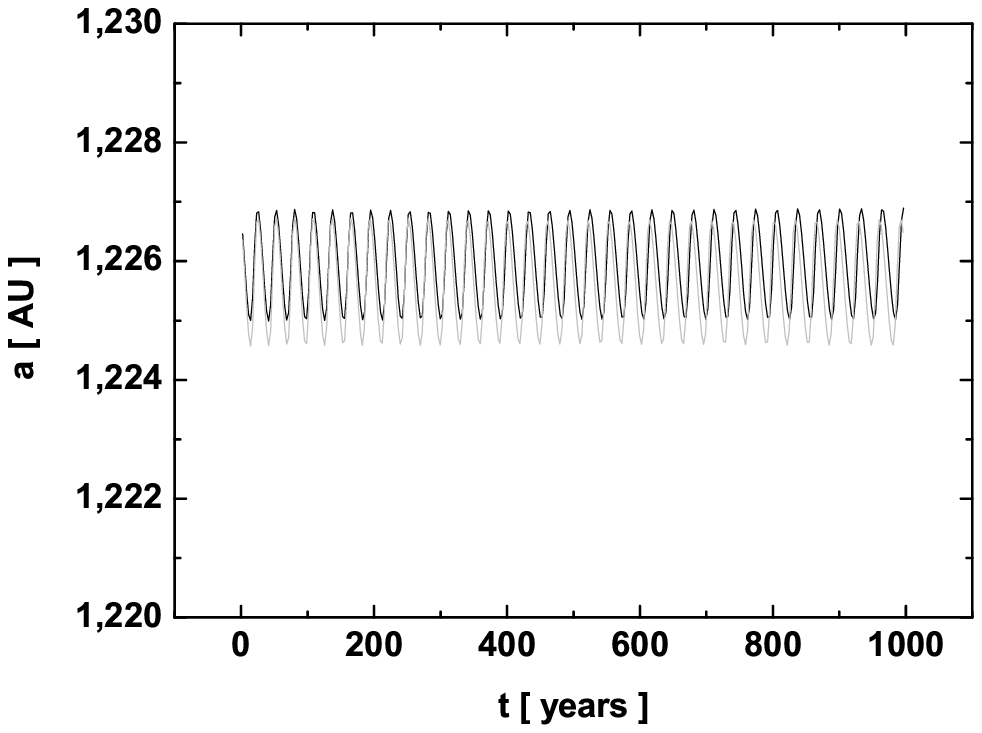}
\end{center}
\end{minipage}
\begin{minipage}[t]{6cm}
\begin{center}
\includegraphics[height=0.20\textheight]{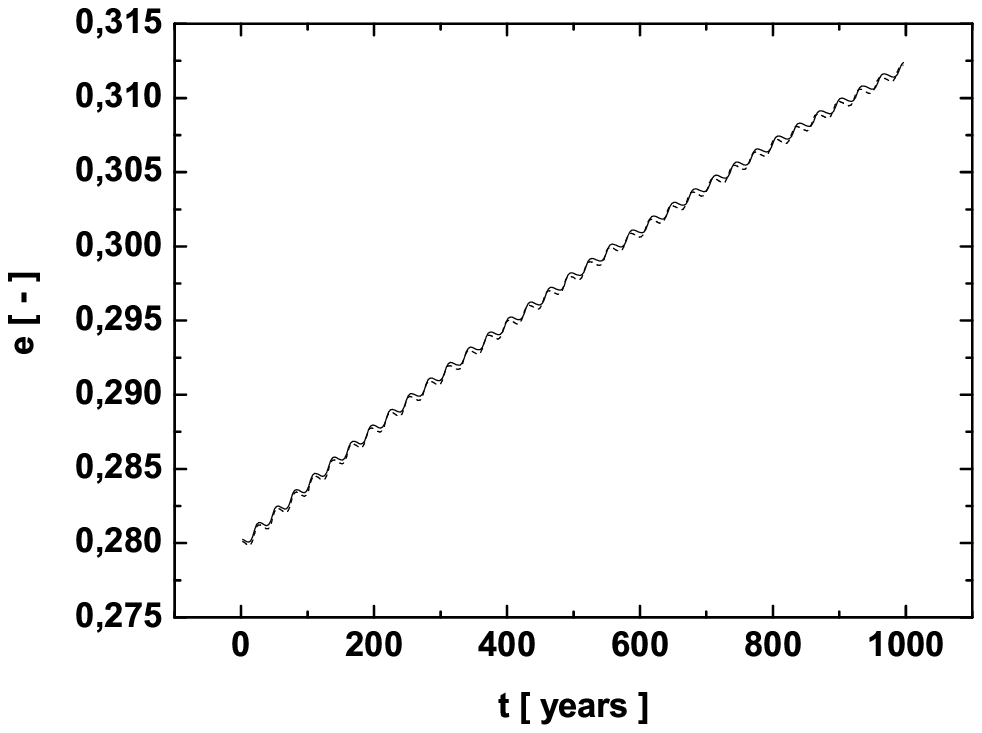}
\end{center}
\end{minipage}
\begin{minipage}[t]{6cm}
\begin{center}
\includegraphics[height=0.20\textheight]{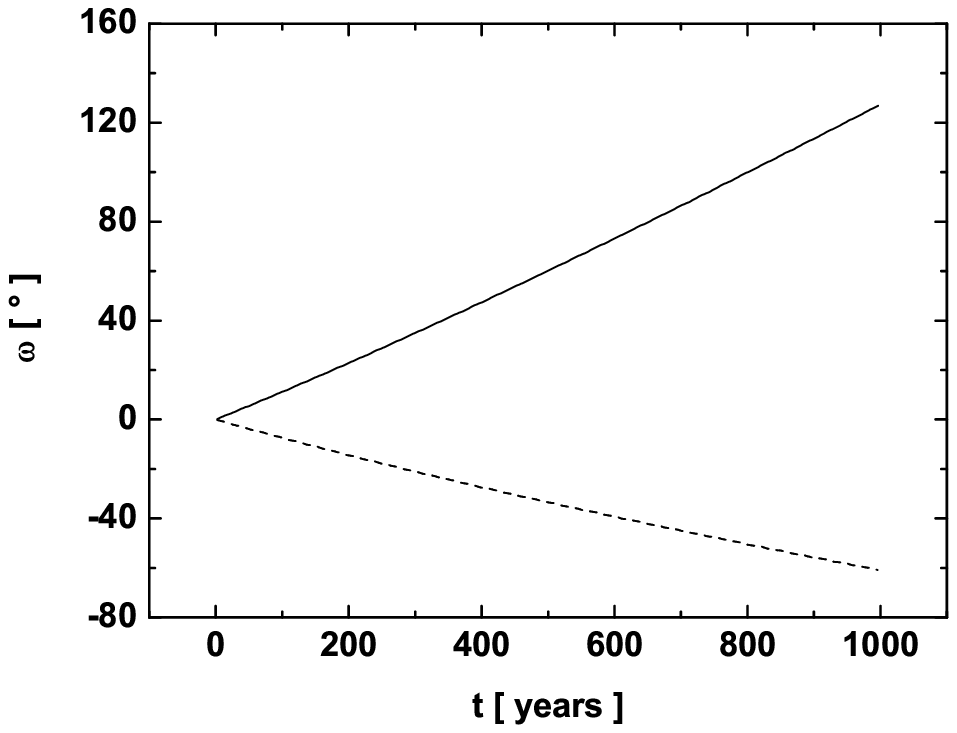}
\end{center}
\end{minipage}
\end{center}
\caption{Black line depict orbital evolution of dust particle with optical
properties dependent on heliocentric distance; radius of the particle $R=2$
$\mu$m, density $\rho=2$ g/cm$^{3}$. Particle is captured in the exterior
resonance $3/2$ with a planet of Earth mass, semi-major axis $a_{P}=1$ AU and
eccentricity $e_{P}=0$. The shift of pericenter is in positive direction. Gray
color or dashed curve represents orbital evolution of the particle in the
resonance with the planet (identical initial conditions), but optical
properties of the particle are constant. The shift of pericenter is in negative
direction for particle with constant optical properties.}
\label{F10}
\end{figure}

It is also possible to find the shift of pericenter in negative direction for
particle with radius $R=2$ $\mu$m and density $\rho=2$ g/cm$^{3}$. This case
is depicted in Fig. 11. However, the shift of pericenter in negative direction
is rare, for this type particle: the situation happens for initial conditions
shown in Fig. 12.

\begin{figure}
\begin{center}
\begin{minipage}[t]{6cm}
\begin{center}
\includegraphics[height=0.20\textheight]{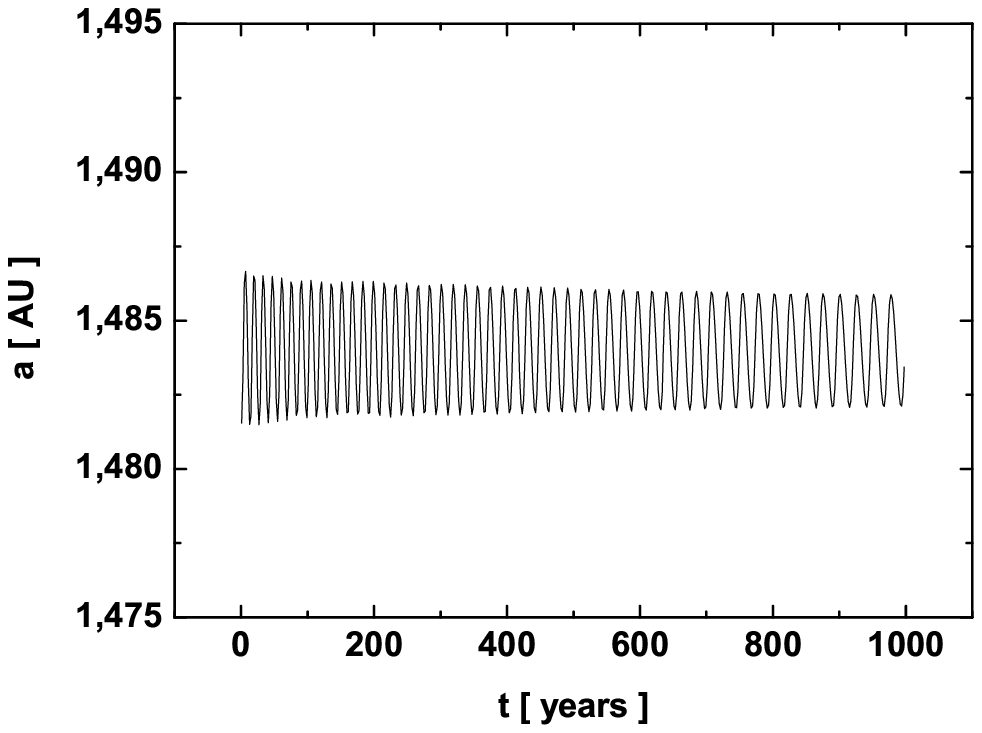}
\end{center}
\end{minipage}
\begin{minipage}[t]{6cm}
\begin{center}
\includegraphics[height=0.20\textheight]{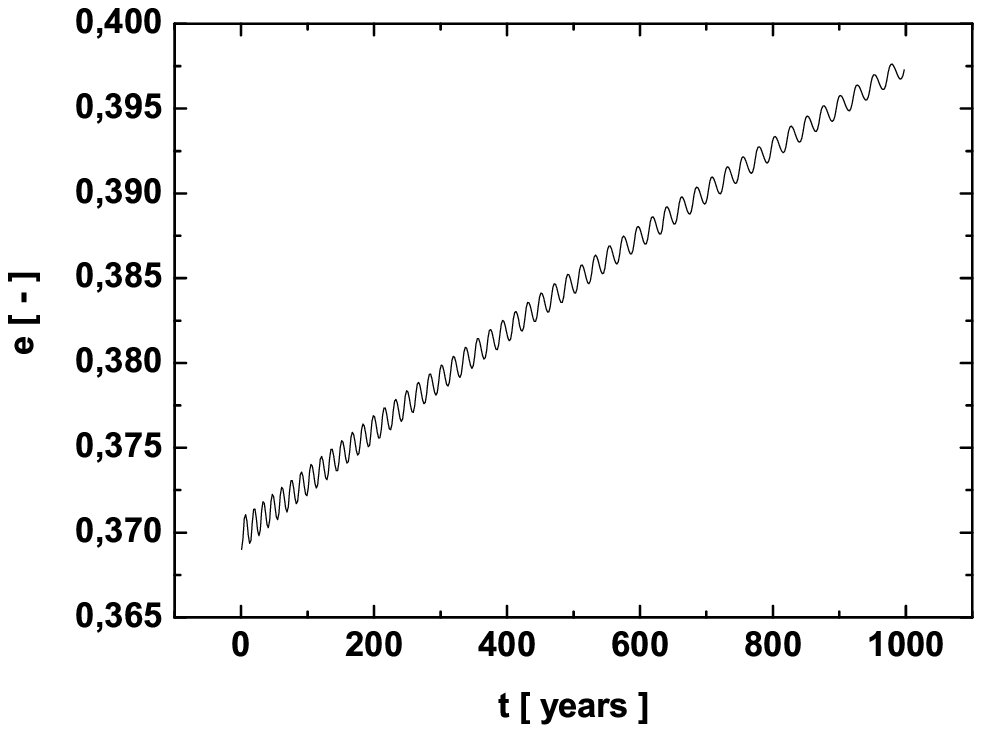}
\end{center}
\end{minipage}
\begin{minipage}[t]{6cm}
\begin{center}
\includegraphics[height=0.20\textheight]{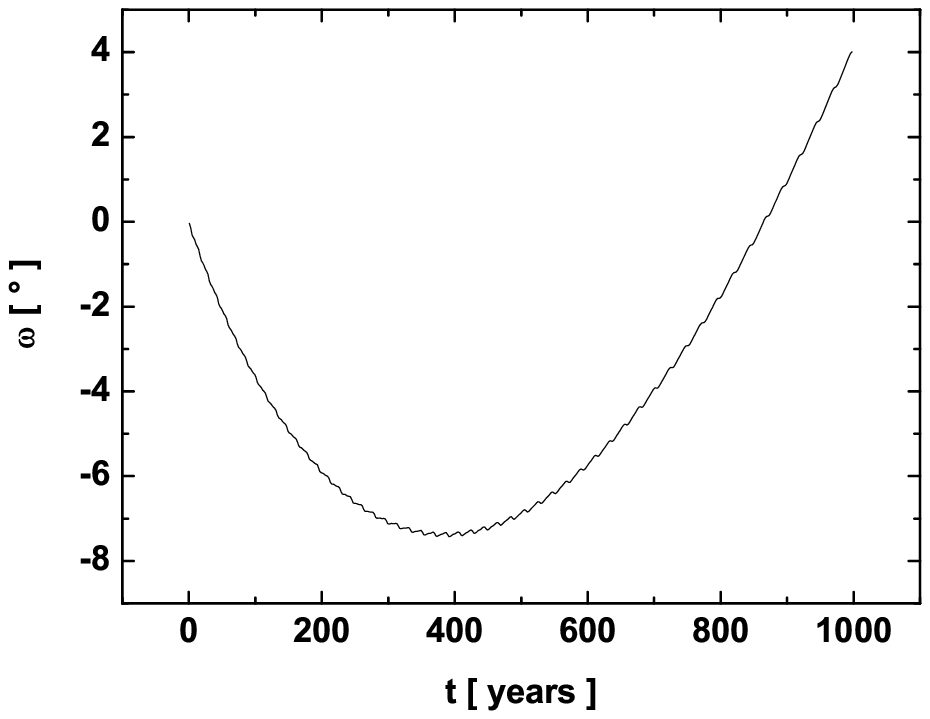}
\end{center}
\end{minipage}
\end{center}
\caption{Evolution of orbital elements of dust particle which optical
properties depend on heliocentric distance. Radius of the particle $R=2$
$\mu$m, density $\rho=2$ g/cm$^{3}$. The particle is captured in the exterior
resonance $2/1$ with a planet of mass equal to the Earth mass, semi-major axis
$a_{P}=1$ AU and eccentricity $e_{P}=0$. Shift of pericenter is in negative
direction for $t\apprle 400$ years. This situation happens for initial
conditions shown in Fig. 12.}
\label{F11}
\end{figure}

\begin{figure}
\begin{center}
\includegraphics[height=0.30\textheight]{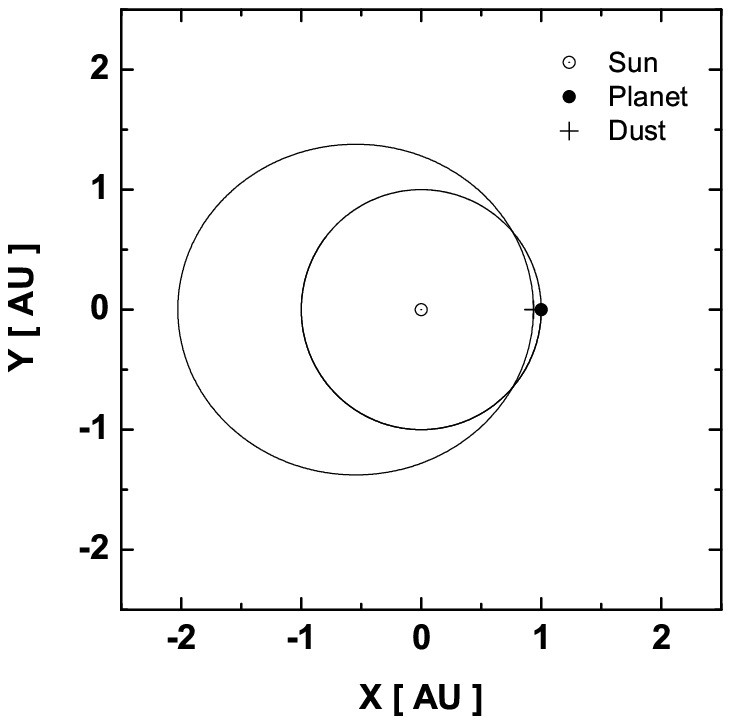}
\end{center}
\caption{Initial conditions for orbital evolution shown in Fig. 11 in the
orbital plane XY. Ellipse represents trajectory of dust particle during the
first two years. Sun is at the origin. Circle represents trajectory of the
planet. Black cross represents the initial conditions of the particle. Black
dot on the right represents initial position of the planet. The particle is
initially inside the planetary orbit, but it is captured in the exterior
resonance.}
\label{F12}
\end{figure}

As for the shift of pericenter, we may use perturbation equations of celestial
mechanics. Central acceleration will be
$-GM_{\odot}(1-\beta_{0})\vec{e}_{R}/r^{2}$, where $\beta_{0}$ is the value of
$\beta$ at time $t=0$. Subtracting the central acceleration from the right-hand
side of Eq. (4) yields the perturbation acceleration. Again, the planar
circular restricted three-body problem and the P-R effect are considered. We
get
\begin{eqnarray}\label{27}
\frac{d\omega_{\beta_{0}}}{dt} &=& -~\frac{1}{e_{\beta_{0}}}~
			\sqrt{\frac{p_{\beta_{0}}}{GM_{\odot}(1-\beta_{0})}}~(\beta-\beta_{0})
			\frac{GM_{\odot}}{r^{2}}~\cos f_{\beta_{0}}+
\nonumber \\
& &    +~\beta ~ \frac{G M_{\odot}}{r^{2}}~
			\frac{1}{c}~\frac{1}{e_{\beta_{0}}}~
			\left (e_{\beta_{0}}\cos f_{\beta_{0}}~-~2 \right )~
			\sin f_{\beta_{0}} ~-~
\nonumber \\
& &    -~\frac{Gm_{P}}{e_{\beta_{0}}}~
			\sqrt{\frac{a_{\beta_{0}}(1-e_{\beta_{0}}^{2})}
			{G M_{\odot} \left (1-\beta_{0} \right )}}\times Y ~,
\nonumber \\
Y &=&  -~\frac{r}{|\vec{r} - \vec{r}_{P}|^{3}}~\cos f_{\beta_{0}}~+~
			\left ( \frac{r_{P}}{|\vec{r}-\vec{r}_{P}|^{3}}-\frac{1}{r_{P}^{2}}
			\right )\times Z ~,
\nonumber \\
Z &=&    \frac{\sin f_{\beta_{0}}}{1+e_{\beta_{0}}\cos f_{\beta_{0}}}
			\sin \left (\Theta_{\beta_{0}}-\Theta_{P} \right )+
			\cos \left (\Theta_{P}-\omega_{\beta_{0}} \right ) ~,
\end{eqnarray}
where $\Theta_{\beta_{0}}=\omega_{\beta_{0}}+f_{\beta_{0}}$ is the position
angle of the particle on the orbit (measured from an X-axis) and $\Theta_{P}$
is the position angle of the planet on the orbit (measured from the X-axis).

Initial conditions depicted in Fig. 12 are: $\beta=\beta_{0}$,
$f_{\beta_{0}}=0^{\circ}$, $\Theta_{\beta_{0}}=0^{\circ}$,
$\omega_{\beta_{0}}=0^{\circ}$, $\Theta_{P} = 0^{\circ}$, and $r$ is a little
less than $r_{P}$. Inserting the initial conditions into Eq. (24), one obtains
$d\omega_{\beta_{0}}/dt<0$ $^{\circ}$/year. If $r$ is a little greater than
$r_{P}$, then $d\omega_{\beta_{0}}/dt>0$ $^{\circ}$/year. When the particle, is
approximately, in the position depicted in Fig. 12, then the first term in
Eq. (27) is negative -- $\beta$ is an increasing function of heliocentric
distance and $\beta_{0}$ is the value of $\beta$ at perihelion of the
particle's orbit ($t = 0$). The sign of the second term depends on the sign of
$\sin f_{\beta_{0}}$ and the third term is negative when the particle's
position corresponds to that depicted in Fig. 12. The third term is dominant
and the shift of pericenter is negative, since $\beta \approx \beta_{0}$,
$f_{\beta_{0}} \approx 0^{\circ}$ and the particle is near the planet. The
oscular evolution of the longitude of pericenter is negative when the particle
is in the interior part of the planetary orbit and the conjunction is near the
particle's perihelion. This effect is relevant also for the secular evolution
of the longitude of pericenter (shift of pericenter): when the particle moves
outside the planetary orbit, then oscular evolution of the longitude of
pericenter may be positive, but averaging over the orbital period yields
negative sign. This may be explanation of Fig. 11 which depicts a decreasing
secular evolution of the longitude of pericenter.

If the motion of dust particle is retrograde and the motion of the planet is
prograde, then the shift of the particle pericenter should be, mainly, in
negative direction, for the changing optical properties of the particle
(Kla\v{c}ka {\it et al.} 2007). However, this depends on the detail behavior of
the change of optical properties.

\section{Conclusions}

The contribution deals with the effect of solar electromagnetic radiation on
dynamics of cosmic dust particles in mean-motion orbital resonances with a
planet of mass equal to the mass of the Earth. We discuss not only the planar
circular restricted three-body problem and the Poynting-Robertson effect with
constant optical properties of the spherical particle. We admit also nonzero
eccentricity of the planet and the change of particle optical properties with
heliocentric distance. The paper concentrates on important properties of motion
of dust grain in the zone of mean-motion resonances. We are interested mainly
in pericenter motion and evolution of eccentricity in the resonances.

Our numerical integrations suggest that any analytic expression for secular
time derivative of the particle's longitude of pericenter does not exist, if a
dependence only on semi-major axis, eccentricity and longitude of pericenter is
considered (P-R effect and mean-motion resonance with planet in circular orbit
are taken into account).

If planetary eccentricity is close to zero and the shift of pericenter is
sufficiently fast, then oscillations of dust grain secular eccentricity exist.
The oscillations occur around the curve corresponding to secular evolution of
the grain eccentricity in the planar circular restricted three-body problem
with the P-R effect. This holds for exterior mean-motion orbital resonances.
Nothing like this was found in the case of interior resonances. However,
interior resonances can exhibit systematic increase of secular eccentricity of
the grain during the capture in the resonances. This is true when eccentricity
of the planet is larger than some critical value depending on the type of the
resonance, mass of the planet and optical parameter $\beta$. The case
$e_{P} = 0$ yields only secular decrease of eccentricity, for interior
resonances.

Our numerical simulations show that noncircularity of the planetary orbit
stabilizes motion of spherical dust grain in the mean-motion orbital
resonances. Maximal capture time (and also mean capture time for many captures
in numerical simulations) of the grain for a given resonance and
particle is greater than it is for the case of circular planetary orbit. This
holds both for exterior and interior resonances. Nonsphericity of the grain
destabilizes motion in the resonances: the greater aspect ratio and smaller
volume of the grain, the shorter capture time.

If a change of optical properties of the spherical grain with heliocentric
distance is also considered, then the shift of pericenter is dominated in
positive direction/orientation for prograde motion of the particle; this holds
both for circular and noncircular planetary orbits and exterior mean-motion
orbital resonances (see, e. g., Fig. 10). If planetary orbit is characterized
by large eccentricity, then secular evolution of dust grain eccentricity may
exhibit complicated behavior.

Spherical dust grain in the planar circular restricted three-body problem
with the Poynting-Robertson effect is characterized by a monotonic secular
evolution of the grain eccentricity (this is true both for constant and
variable optical properties of the grain), in the exterior mean-motion orbital
resonances. Such kind of behavior does not exist if at least one of the
above mentioned assumptions -- sphericity of the grain or circular orbit of the
planet -- is cancelled. As for the interior resonances, the circular planetary
orbit yields secular decrease of the eccentricity, while nonsphericity of the
grain or noncircularity of the planetary orbit may yield secular increase of
the eccentricity, also. There is some kind of unification of qualitative
kinematical behavior of interplanetary dust grains under the action of more
real physical forces: there is not great difference between evolution of
orbital elements of spherical and nonspherical dust grains if more general
physical forces are taken into account. Our results show that gravitational
effect can mimic nongravitational effect. However, the following question is
still unanswered: Can spherical grain be captured into a resonance when the
secular evolution of particle semi-major axis is an increasing function of
time?

\acknowledgements{}
The paper was supported by the Scientific Grant Agency VEGA
(grant No. 1/3074/06).

\end{document}